\documentclass[aps,pre,twocolumn,amsmath,amssymb,nofootinbib,superscriptaddress,showpacs,longbibliography]{revtex4-1}

\usepackage{mathrsfs}
\usepackage{graphicx}
\usepackage{natbib}
\usepackage{latexsym}
\usepackage{amsmath, amssymb, amsthm}
\usepackage{bbold}
\usepackage{mathtools}
\usepackage{xcolor}
\usepackage{comment}
\usepackage{wasysym}

\newcommand*\Laplace{\mathop{}\!\mathbin\bigtriangleup}

\newcommand{\change}[1]{{\color{black}{#1}}}

\newcommand{\av}[1]{\langle #1 \rangle}
\newcommand{\km}{k_{\text{min}}}
\newcommand{\kmax}{k_{\text{max}}}
\newcommand{\be}{\begin{equation}}
\newcommand{\ee}{\end{equation}}

\usepackage[bottom]{footmisc}

\usepackage{microtype}
\usepackage{epigraph}
\usepackage{lipsum}

\usepackage{float}

\usepackage[normalem]{ulem}

\usepackage{appendix}

\usepackage{hyperref}

\begin{document}

\title{Scaling and universality for percolation in random networks: a unified view}

\author{Lorenzo Cirigliano}
\affiliation{Dipartimento di Fisica Universit\`a di Roma ``\href{https://ror.org/02be6w209}{La Sapienza}”, P.le
  A. Moro, 2, I-00185 Rome, Italy}

\affiliation{\href{https://ror.org/01qb1sw63}{Centro Ricerche Enrico Fermi}, Piazza del Viminale, 1,
  I-00184 Rome, Italy}

\author{G\'abor Tim\'ar}
\affiliation{School of Mathematics, \href{https://ror.org/024mrxd33}{University of Leeds}, Leeds LS2 9JT, UK}

\author{Claudio Castellano}
\affiliation{\href{https://ror.org/05rcgef49}{Istituto dei Sistemi Complessi (ISC-CNR)}, Via dei Taurini 19, I-00185 Rome, Italy}

\affiliation{\href{https://ror.org/01qb1sw63}{Centro Ricerche Enrico Fermi}, Piazza del Viminale, 1,
  I-00184 Rome, Italy}

\date{\today}

\begin{abstract}
Percolation processes on random networks have been the subject
of intense research activity over the last decades: the overall
phenomenology of standard percolation on uncorrelated and unclustered
topologies is well known. Still some critical properties of the
transition, in particular for heterogeneous substrates, have not
been fully elucidated and contradictory results appear in
the literature. In this paper we present, by means of a generating
functions approach, a thorough and complete investigation of percolation
critical properties in \change{uncorrelated locally tree-like} random networks.
We determine all critical exponents, the associated critical
amplitude ratios and the form of the cluster size distribution
for networks of any level of heterogeneity.
We uncover, in particular for highly heterogeneous networks,
subtle crossover phenomena, nontrivial scaling forms and
violations of hyperscaling.
In this way we clarify the origin of inconsistencies in the previous literature.

\end{abstract}


\maketitle

\section{Introduction}
\label{sec:introduction}
Percolation is one of the simplest and
most-studied processes in statistical physics. It consists in damaging
a graph $\mathcal{G}=(\mathcal{V},\mathcal{E})$, with
$N=|\mathcal{V}|$ nodes and $E=|\mathcal{E}|$ edges, by removing some
of the edges in $\mathcal{E}$. This damaging is typically pursued
either by deactivating nodes and removing from $\mathcal{E}$ all the
edges belonging to deactivated nodes -- this is called \textit{site
  percolation} -- or by directly removing the edges from $\mathcal{E}$
-- this is called \textit{bond percolation}. Percolation theory
studies the properties of the graph's connected components, or
clusters -- maximal subsets of $\mathcal{V}$ in which there is at least one path
joining each pair of nodes -- after the damage. Owing to its
generality, percolation is used to model a wide variety
of phenomena, e.g., forest fires \cite{stauffer2018introduction},
porous media \cite{machta1991phase, moon1995critical},
ionic transport in composite materials \cite{sahimi1994applications},
epidemic spreading \cite{grassberger1983critical, cardy1985epidemic, newman2002spread},
network robustness \cite{cohen2000resilience, callaway2000network, holme2002attack},
and it constitutes a paradigmatic example of systems undergoing phase transitions.

If all nodes \change{(edges)} are removed, the graph does not contain any
connected component \change{(contains only components of size $1$)}.
On the other hand, if the fraction of removed nodes or
edges is small, in a very large system an extensive connected component of size
$\mathcal{S} \sim N$ \change{is usually present,
except for specific graphs, e.g. one-dimensional lattices}.
In the thermodynamic limit $N \to \infty$, this cluster is called
giant component (GC), while all other clusters are small, i.e., nonextensive.
The shift from a phase without a giant component (non-percolating
phase) to a phase in which a giant component exists (percolating
phase) is the percolation transition, and it typically involves the
same kind of singularities of systems undergoing continuous phase
transitions~\cite{goldenfeld2018lectures,patashinsky1979fluctuation}:
the critical behavior observed in different models can be described
with a set of universal critical exponents, shared by models having
different microscopic details but the same symmetries.

Percolation on
``finite-dimensional'' topologies, such as \change{$D$}-dimensional regular
lattices, is supported by a robust theoretical background, whose roots
can be traced back to renormalization group
theory~\cite{stauffer2018introduction,christensen2005complexity}.
\change{The exact values of the critical exponents are known for $D=2$ \cite{dennijs1979relation,nienhuis1980magnetic,blote1981critical,smirnov2001critical}, and for $D \geq 6$ the system is described by the mean-field critical exponents \cite{aharony1980universal}.} \change{Even if exact solutions are in general not available for arbitrary $D$},
scaling relations among the critical exponents and perturbative approaches
provide a full understanding and precise numerical
evaluations~\cite{aharony1980universal}.

For percolation in complex networks, the situation is slightly
different. \change{Starting from the seminal works on random graphs by P. Erd\H{o}s and A. R\'enyi \cite{erdos1959random,erdos1960evolution},} several exact results \change{have been obtained} for percolation on
various complex topologies, from \change{uncorrelated random
graphs~\cite{molloy1995critical,molloy1998size,joseph2014component,karrer2014percolation,hamilton2014tight}}
to random graphs with degree-degree correlations~\cite{goltsev2008percolation},
to some network models with many short loops \cite{cantwell2019message}.
Other works~\cite{cohen2002percolation,dorogovtsev2008critical,li2021percolation}
have provided a general heuristic picture of the percolation phase-transition
in complex networks, that is generally believed to be complete and
coherent.
However, a close scrutiny reveals that several results in the literature
are contradictory, while other aspects of the phenomenology have not
been fully clarified.
An example of this confusion concerns the value of the Fisher exponent $\tau$
(see below for a precise definition) for strongly heterogeneous
networks (with degree distribution $p_k \sim k^{-\gamma_d}$ and $2<\gamma_d<3$).
In this range of $\gamma_d$ values $\tau$ was claimed to be equal
to $3$~\cite{dorogovtsev2008critical}, to
$2+1/(\gamma_d-2)$~\cite{cohen2002percolation}, to
$\gamma_d$~\cite{lee2004evolution} or even to be not
defined~\cite{radicchi2015breaking}.
Similarly, contradictory values are claimed for the exponent $\bar \nu$
\change{(sometimes misleadingly indicated as
$\nu$~\cite{radicchi2015breaking,li2021percolation})} governing finite size scaling:
$2/(3-\gamma_d)$~\cite{dorogovtsev2008critical},
$(\gamma_d-1)/(3-\gamma_d)$~\cite{cohen2002percolation,li2021percolation}.
In addition, the scaling property of the cluster size distribution $n_s$
has been postulated but the explicit form of the scaling function in
all the ranges of $\gamma_d$ values has not been determined, although
exact results for the component size distribution of uncorrelated
random graphs have recently been published~\cite{newman2007component,kryven2017general}.
Finally, to the best of our knowledge, universal amplitude ratios for
percolation in networks have not been computed so far.
In this manuscript our goal is to provide a complete and coherent
summary of all critical properties of the percolation
phase-transition in \change{uncorrelated locally tree-like} random networks, that can be used for reference.
To achieve this objective we fully exploit the power of the generating
functions approach, which was only partly used before, to determine
all critical exponents and scaling functions.
In this way we point out and clarify all the inconsistencies present
in the literature, derive all the missing results and
independently test the validity of scaling relations.

The rest of the paper is organized as follows.
Sec.~\ref{sec:scaling} sets the stage, by means of a very general presentation of the
percolation problem on networks: we define the model, the 
observables of interest and their critical properties.
We then present the scaling ansatz for the cluster size
distribution, the scaling relations among critical exponents,
the definition of the correlation length and of the finite-size
scaling properties.
This content is not original but a uniform and complete presentation
is crucial to make the more specific treatment of the other sections
easy to follow.
In Section~\ref{sec:recipe} we consider an ensemble of \change{uncorrelated locally tree-like} random networks
and determine closed expressions for all relevant critical
properties in terms of generating functions.
In Section~\ref{results} we summarize the main results of the manuscript, i.e.
a quantitative description of the percolation
critical properties for homogeneous and for
power-law degree-distributed random networks.
For the sake of readability
the explicit determination of these results, obtained by
carefully applying the theory of Section~\ref{sec:recipe},
is deferred to an Appendix.
Some of the analytical results are tested via numerical simulations
in Section~\ref{simul}.
A summary of the main findings and a discussion of their
implications concludes the manuscript in Section~\ref{sec:conclusions}.

\section{Percolation on networks, scaling and critical properties}
\label{sec:scaling}
In this Section, we define the main quantities of interest in
percolation and we outline the differences between site and bond
percolation. We then introduce a general scaling ansatz for critical
properties and discuss its consequences.

\subsection{Site and bond percolation}
Let us consider first the case of uniform site percolation, in which
every node is active with probability $\phi$ -- i.e., inactive with
probability $1-\phi$. The activation probability $\phi$ plays the role
of control parameter.
The order parameter of percolation is usually defined as the relative size
of the GC, averaged over the randomness of the percolation process,
$P^{\infty}(\phi)=\lim_{N \to \infty} \langle \mathcal{S} \rangle/N$.
$P^{\infty}(\phi)$ is also the probability that a randomly chosen node
belongs to the GC. Then the percolation transition separates a
phase in which $P^{\infty}=0$, for $\phi \le \phi_c$,
from a phase with $P^{\infty}>0$, for
$\phi > \phi_c$; $\phi_c$ is the percolation threshold.

The basic quantity in percolation theory is the cluster size distribution,
defined as the number of small clusters of size $s$ per active node,
i.e., for site percolation,
\begin{equation}
  n_s(\phi) \equiv \frac{{\cal N}_s(\phi)}{\phi N},
  \label{ns}
\end{equation}
where ${\cal N}_s(\phi)$ is the number of clusters of size $s$.
Note that the largest cluster is excluded from the enumeration of clusters
in $n_s(\phi)$.
A cluster size distribution including also the largest cluster of size $\mathcal{S}$ can be defined,
\begin{equation}
\widetilde{n}_s(\phi)=n_s(\phi)+\frac{\delta_{s, \mathcal{S}}}{\phi N},
\label{eq:component_with_largest}
\end{equation}
where $\mathcal{S}$ is a random variable distributed according to $\mathcal{P}(s=\mathcal{S},\phi)$. Away from criticality, $\mathcal{P}(s=\mathcal{S},\phi)$
is a Gaussian centered in $\langle \mathcal{S} \rangle$ with fluctuations
of order $\sqrt{N}$.
In the limit of diverging $N$ we can thus write
\begin{equation}
  \frac{\delta_{s, \mathcal{S}}}{\phi N} \to \frac{\delta_{s,\av{\mathcal{S}}}}{\phi N}
  = \frac{\delta_{s,\av{\mathcal{S}}}}{\phi \av{\mathcal{S}}} P^\infty(\phi).
  \label{PS}
\end{equation}
In this same limit, summing Eq.~\eqref{eq:component_with_largest}
over $s$ yields
$\sum_s \widetilde{n}_s(\phi)=\sum_s n_s(\phi)$, because $\av{\mathcal{S}}$
in the
denominator diverges.
The two distributions are not normalized.

The normalized distribution is instead $s \widetilde{n}_s(\phi)$,
which is the probability that a randomly chosen active node belongs to a
cluster of any size $s$. Its normalization simply reflects the fact that
any active node must be part of a cluster.
Multiplying Eq.~\eqref{eq:component_with_largest} by $s$, using
Eq.~\eqref{PS} and summing over $s$ we obtain
\begin{equation}
1 = \sum_{s \geq 1} s n_s(\phi) + \frac{P^{\infty}(\phi)}{\phi}.  
\end{equation}

The distribution $\pi_s(\phi) \equiv s n_s(\phi)$ is normalized
only when there is no giant component, $P^{\infty}=0$. Otherwise
\begin{equation}
 P^{\infty}(\phi) = \phi\left[ 1-\sum_{s \geq 1}\pi_s(\phi) \right].
 \label{eq:order_parameter}
\end{equation}
This equation connects the singular behavior observed in the order
parameter at $\phi=\phi_c$ with $\pi_s(\phi_c)=sn_s(\phi_c)$.

The same argument developed so far holds also for bond
percolation, in which edges are kept with probability $\varphi$, and
all nodes are considered to be active, which implies division
by $N$ (instead of $\phi N$) in Eqs.~\eqref{ns} and~\eqref{PS}.
The theory in this paper is formally developed for site percolation,
however, the same theory applies to bond percolation as well, provided
Eq.~\eqref{eq:order_parameter} is replaced
with
\begin{equation}
 P^{\infty}_{(\text{b})}(\varphi)=1-\sum_{s \geq 1}\pi^{(\text{b})}_s(\varphi).
 \label{eq:order_parameter_bond}
\end{equation}

It is clear that $P^{\infty}(\phi)=\phi P^{\infty}_{(\text{b})}(\phi)$
if and only if $\pi_s(\phi) = \pi_s^{(\text{b})}(\phi)$, that is if
activating a fraction $\phi$ of the nodes or a fraction $\phi$ of the
edges produces exactly the same effect on the cluster size
distribution. Throughout the paper we will comment on relevant
differences between the behavior of site and bond percolation.

\subsection{Observables and critical properties}
The behavior of the system is characterized by considering the
following observables.

\begin{equation}
  F(\phi) \equiv \sum_{s=1}^\infty n_s(\phi)
  \label{F}
\end{equation}
is the mean number (per active node) of finite clusters in the system.
When the control parameter $\phi$ approaches the critical threshold,
the singular part of this quantity scales as
\begin{equation}
F(t) \simeq A_{\pm} |\phi-\phi_c|^{2-\alpha} = A_{\pm} |t|^{2-\alpha},
\end{equation}
where $\alpha$ is a universal critical exponent and
$t=\phi-\phi_c$\footnote{Here and in the rest of the paper we use,
with a slight abuse of notation, the same symbols to denote functions
of $\phi$ and functions of $t$.}

The relative size of the giant component, or percolation strength,
is defined in Eq.~\eqref{eq:order_parameter} (for site percolation)
and in Eq.~\eqref{eq:order_parameter_bond} (for bond percolation).
This quantity is zero in the nonpercolating phase, while it is finite
above the threshold. Close to criticality it behaves as
\begin{equation}
P^{\infty}(t) \simeq  B t^{\beta}
\label{eq:P_infty_def}
\end{equation}
where $\beta$ is another critical
exponent. This exponent differs between site and bond percolation when $\phi_c=0$, because of the multiplicative factor $\phi$ present
in Eq.~\eqref{eq:order_parameter}:
$\beta_{site} = 1 + \beta_{bond}$~\cite{radicchi2015breaking}.
Apart from this case the two exponents always coincide and to avoid ambiguity we define
as $\beta$ the exponent associated to the singular behavior of the
factor $1-\sum_s \pi_s$ present in both definitions,
Eq.~\eqref{eq:order_parameter} and Eq.~\eqref{eq:order_parameter_bond}.

Another observable is the mean size of the cluster to which
a randomly chosen \change{active} node (not in the giant component) belongs, which is
\begin{equation}
  \langle s \rangle = 
  \frac{\sum_{s} s^2 n_s(\phi)}{\sum_{s} s n_s(\phi)}.
\end{equation}
At criticality this quantity has a singular behavior, with an
associated critical exponent $\gamma$
\begin{equation}
\langle s \rangle -1 \simeq C_{\pm} |t|^{-\gamma}.
\label{eq:average_s_def}
\end{equation}

The goal of a theory for percolation is the determination of the universal
critical exponents and of the universal critical amplitude ratios.
Following Ref. \cite{aharony1980universal}, we introduce the so called ``ghost
field'' $h$ and generalize Eq.~\eqref{F} defining
\begin{align}
\Omega(\phi, h)&\equiv \sum_{s=1}^{\infty}n_s(\phi)e^{-sh}.
\label{eq:omega_def}
\end{align}

The observables defined above can be expressed in terms of
$\Omega$ and of the derivatives of $\Omega$ with respect to $h$.
Clearly $F(\phi)=\Omega(\phi,0)$.
We further define
\begin{equation}
  \Psi(\phi, h) \equiv -\frac{\partial \Omega}{\partial h} =
  \sum_{s=1}^{\infty} s n_s(\phi)e^{-sh},
\end{equation}
which is related to the site percolation order parameter
\begin{equation}
  P^{\infty}(\phi)=\phi \big[1-\Psi(\phi,0)\big]
  \label{eqP}
\end{equation}
and similarly for bond percolation, without the global factor $\phi$,
see Eq.~\eqref{eq:order_parameter_bond}.
Finally, the quantity
\begin{equation}
  \chi(\phi, h) \equiv \frac{\partial^2 \Omega}{\partial h^2} = \sum_{s=1}^{\infty}s^2 n_s(\phi)e^{-sh},
\end{equation}
is related to $\langle s \rangle$ since,
\begin{equation}
  \langle s \rangle =
  \frac{\sum_{s} s^2 n_s(\phi)}{\sum_{s} s n_s(\phi)} 
  = \frac{\chi(\phi,0)}{\Psi(\phi,0)}
  \label{s}.
\end{equation}
Note that, using $n_s(\phi)=\pi_s(\phi)/s$ and
$s^{-1}=\int_{0}^\infty dh e^{-hs}$, $F(\phi)$ can be written as
\begin{equation}
 F(\phi)=\int_{0}^{\infty}dh \sum_{s \geq 1} \pi_s(\phi)e^{-sh}=\int_{0}^{\infty}dh \Psi(\phi,h).
 \label{eq:N_t_integral}
\end{equation}

In the critical region near $\phi_c$, setting $t=\phi-\phi_c$,
we can write
\change{
\begin{equation}
 \Omega(t,h) = \sum_{n=0}^{\infty}a_n(t) h ^n + \left \{\Omega(t, h)\right \}_{\text{sing}},
 \label{eq:omega_regular_singular}
\end{equation} }
where $\{\cdot \}_{\text{sing}}$ stands for ``singular part''.
\change{Note that Eq.~\eqref{eq:omega_regular_singular} is completely
  general. Indeed, we can write $n_s(t) = n_s^{\text{small}}(t) + n_s^{\text{tail}}(t)$,
  where $n_s^{\text{small}}$ decays fast enough for $s \gg 1$,
  and $n_s^{\text{tail}}$ is the large $s$ tail of the
  distribution. Using Eq.~\eqref{eq:omega_def}, expanding $e^{-sh}$ in
  power series and changing the order of summation, the contribution
  from $n_s^{\text{small}}$ gives
\begin{equation*}
  \sum_{s=1}^{\infty}n_s^{\text{small}}(t) e^{-sh} = \sum_{n=0}^{\infty} \left[(-1)^{n}\frac{\sum_{s=1}^{\infty}s^n
      n_s^{\text{small}}(t)}{n!} \right]h^n,
\end{equation*}
which is the first, regular, term on the r.h.s. of
Eq.~\eqref{eq:omega_regular_singular}. This procedure is not
mathematically justified a priori for $n_s^{\text{tail}}$ because the series
in the square brackets may diverge, if $n_s^{\text{tail}}$ decays
slowly. Hence the large $s$ tail of the cluster size distribution is
the origin of singular terms in
Eq.~\eqref{eq:omega_regular_singular}.}  The critical behavior is due
to the singular part of $\Omega(t,h)$ when the point $(t=0,h=0)$ is
approached.  In addition to the critical exponents defined above
another exponent can be defined as
\begin{align}
1-\Psi(\phi_c,h) &\simeq E h^{1/\delta},
\label{eq:definition_delta}
\end{align}
where in the last equation logarithmic corrections may appear for integer
$1/\delta$.

In summary, the critical behavior in terms of
$\{\Omega(t,h)\}_{\text{sing}} $ and its derivatives is, for $|t| \ll
1$ and $h \ll 1$,

\begin{align}
\label{eq:scaling_exponent_1}
\Omega(t, 0) &\simeq A_{\pm} |t|^{2-\alpha},\\
\label{eq:scaling_exponent_2}
1-\Psi(t, 0) &\simeq B t ^{\beta},\\
\label{eq:scaling_exponent_3}
\frac{\chi(t, 0)}{\Psi(t,0)}-1 & \simeq C_{\pm}|t|^{-\gamma},\\
\label{eq:scaling_exponent_4}
1-\Psi(0,h) &\simeq E h^{1/\delta}.
\end{align}

\subsection{The scaling ansatz for the cluster size distribution}

At its very core, the scaling hypothesis for percolation can be
reduced to the following ansatz for the \change{tail of the }cluster size
distribution~\change{\cite{stauffer1975violation,stauffer1979scaling}}
\begin{equation}
n_s(t) \simeq q_0 s^{-\tau}f_{\pm}(q_1 t s^{\sigma}),
\label{eq:scaling}
\end{equation}
for $s \gg 1$ and $|t| \ll 1$, where $q_0, q_1$ are nonuniversal positive
constants, $f_{\pm}(x)$ are universal scaling functions, while
$\tau$ and $\sigma$ are universal critical exponents.
The scaling functions decay to zero quickly for large values of their argument:
$f_{\pm}(x) \ll 1$ for $|x| \gg 1$.
This naturally introduces the quantity
\begin{equation}
 s_{\xi}\simeq \Sigma |t|^{-1/\sigma},
 \label{eq:definition_correlation_size}
\end{equation}
which plays the role of a correlation size, where $\Sigma=q_1^{-1/\sigma}$.
For $s/s_\xi \ll 1$, the cluster size distribution scales as
$n_s(t)\sim s^{-\tau}$.
This implies, if $\sigma >0$ so that
$s_\xi \to \infty$ as $t \to 0$, that $n_s$ exhibits a pure power-law tail for large $s$
with exponent $\tau$.

Note that in the literature the scaling form Eq.~\eqref{eq:scaling}
is often assumed~\cite{newman2001random,newman2018networks,cohen2002percolation} to
be $s^{-\tau} \exp(-s/s_\xi)$,
which corresponds to a scaling function of the form
$f_\pm(x)=e^{-x^{1/\sigma}}$.
As it will be shown below, it is instead crucial to allow
for more general functional forms.

For some particular classes of random networks, it is possible to explicitly calculate
the cluster size distribution $n_s(t)$, finding an agreement with the
general form, Eq.~\eqref{eq:scaling}. Some examples are worked out in
Appendix~\ref{AppendixA}.

The scaling ansatz implies the existence of some scaling relations
among the critical exponents.
From the scaling ansatz,
Eq.~\eqref{eq:scaling}, by writing $\{\Omega(t,h)\}_{\text{sing}}
\simeq q_0 \int_{1}^{\infty}ds s^{-\tau}f_{\pm}(q_1 ts^{\sigma})
e^{-sh}$ and using the change of variable $x=q_1|t|s^{\sigma}$, we
have
\[\{\Omega(t,h)\}_{\text{sing}}  \sim |t|^{\frac{\tau-1}{\sigma}} \int_{q_1 t}^{\infty} dx x^{\frac{1-\tau}{\sigma}-1}f_{\pm}(x) e^{-\left[x^{\frac{1}{\sigma}}h/(q_1|t|)^{1/\sigma}\right]}.\]
The integral depends on $t$ because of the lower extremum, and through the ratio $h/|t|^{1/\sigma}$. The integral is divergent for $t\to 0$. However, because of the multiplicative factor $|t|^{\frac{\tau-1}{\sigma}}$ this divergence is removed.
It follows that $\{\Omega(t,h)\}_{\text{sing}}$ takes the form
\begin{equation}
  \{\Omega(t,h)\}_{\text{sing}}  = |t|^{\frac{\tau-1}{\sigma}} \mathcal{F}_{\pm}(h/|t|^{1/\sigma}),
  \label{eq:Omega_scaling_form}
\end{equation}
where $\mathcal{F}_{\pm}(x)$ are scaling functions such that
$\mathcal{F}_{\pm}(x) \sim 1$ for $|x| \ll 1$. This directly follows
from simple assumptions on the behavior of $f_{\pm}(x)$ for small
$|x|$.\footnote{It is sufficient to assume that $f(x) \sim x^a $ for
  $x\ll1$, $a \geq 0$, including the standard case in which $f(x\ll1)$
  is a constant, i.e., $a=0$.}

Comparing Eq.~\eqref{eq:Omega_scaling_form} with
Eq.~\eqref{eq:scaling_exponent_1} yields $2-\alpha=(\tau-1)/\sigma$.
Taking derivatives of Eq.~\eqref{eq:Omega_scaling_form} with respect to
$h$, evaluated at $h=0$ and comparing with
Eqs.~\eqref{eq:scaling_exponent_2}-\eqref{eq:scaling_exponent_3}
gives two other scaling relations among critical exponents:
$\beta=(\tau-2)/\sigma$ and $\gamma=(3-\tau)/\sigma$.  Hence all
the exponents introduced so far can be expressed as functions of only two of them, for instance
\begin{align}
 \alpha&= 2-\frac{\tau-1}{\sigma},\\
 \gamma &= \frac{3-\tau}{\sigma},\\
\beta&= \frac{\tau-2}{\sigma},\\
\delta &= \frac{1}{\tau-2}.
  \label{scalingrels}
\end{align}

The last of the scaling relations in Eq.~\eqref{scalingrels} is derived 
by means of the so-called Tauberian theorems~\cite{Flajolet2009}.
In particular, for a function $w(h)=\sum_{s\geq 0 }a_s e^{-sh}$,
we have\footnote{There are logarithmic corrections for integer $1/\delta$.}
\begin{equation}
  a_s \simeq E' s^{-1/\delta-1} \Longleftrightarrow w(h) \simeq 1-Eh^{1/\delta}.
\end{equation}
This implies that, since at criticality
$\Psi(\phi_c,h) = \sum_{s \geq 0} s^{1-\tau} e^{-sh}$,
then $\delta = 1/(\tau-2)$.
Note that this relation holds only when $\Psi(\phi_c,h)$ is nonanalytical in $h$.
As shown below, there are cases in which
$\Psi(\phi_c,h)$ is analytical in $h=0$; in these cases the scaling relation
between $\delta$ and $\tau$ breaks down.

The framework developed here is fully general
and it is applicable provided the scaling ansatz Eq.~\eqref{eq:scaling} is valid.

\subsection{The correlation length}

In a finite $D$-dimensional  lattice,
the probability that one node $j$ in $\vec{y}$, at euclidean distance $r_{ij}$ from
an active node $i$ in $\vec{x}$, is in the same connected component as $i$ 
is called the pair correlation function $g(i,j)$.
In particular, we can write $g(i,j)=\langle c_{ij} \rangle$, where
$c_{ij}=1$ if there is a path connecting $i$ and $j$,
and it is zero otherwise, and the average is taken over
the randomness of the percolation process (and over the network ensemble, in the case of random networks).
An important sum rule for the correlation function is
\begin{equation}
 \frac{1}{\phi N}\sum_{i,j \in \mathcal{C}}g(i,j) = \chi(\phi,0),
 \label{eq:sum_rule_correlation}
\end{equation}
where $\mathcal{C}$ denotes the set of nodes belonging to small components. This can be seen from
\begin{equation}
  \frac{1}{\phi N}\sum_{i,j \in \mathcal{C}}\langle c_{ij} \rangle = \frac{1}{\phi N}\sum_{ \mathcal{C}} \langle s_{\mathcal{C}}^2 \rangle = \sum_{s\geq 1}\frac{\mathcal{N}_s(\phi)}{\phi N} s^2,
  \label{eq:sum_rule_correlation2}
\end{equation}
and Eq.~\eqref{eq:sum_rule_correlation} follows since
$n_s(\phi)=\mathcal{N}_s(\phi)/(\phi N)$ by definition.

Using $g(i,j)$, one defines a (euclidean) correlation length $\xi$ via
\begin{equation}
 \xi^2 =\frac{\sum_{i,j \in \mathcal{C}}r^2_{ij}g(i,j)}{\sum_{i,j \in \mathcal{C}}g(i,j)}.
\end{equation}
Note that if $g(i,j)$ depends only on $r_{ij}$, summing over
distances instead of summing over pairs implies
\[\xi^2 = \frac{\sum_{r} r^2 \widetilde{\mu}(r)}{\sum_{r }\widetilde{\mu}(r)}, \]
where $\widetilde{\mu}(r)$ denotes the average number of nodes that are at a distance $r$ from a given node in the same \change{finite} component\change{, see \cite{Cohen2004}}.

In the case of percolation on networks, we
can embed a generic finite tree-like \change{component}
in a \change{infinite}-dimensional lattice by
placing one node in the origin and placing each subsequent neighboring
node along a new orthogonal direction.
Hence, the length $l$ of the shortest path connecting two nodes,
allows us to define the euclidean distance between these two nodes as $r=\sqrt{l}$~\cite{wattendorf2024sublattice}.
Then a correlation length for complex networks can be computed via
\begin{equation}
\xi^2 = \frac{\sum_{l} l \mu(l)}{\sum_{l}\mu(l)} =
\frac{\sum_{l} l \mu(l)}{\chi(\phi,0)},
\label{xi2nets}
\end{equation}
where $\mu(l)$ is the average number of active nodes at
distance $l$ from a given active node in the same \change{finite} component
and Eq.~\eqref{eq:sum_rule_correlation} has been used.

In all cases, close to the critical point,
\begin{equation}
  \xi \simeq \Xi_{\pm}|t|^{-\nu}.
  \label{xi}
\end{equation}

This defines a new critical exponent $\nu$.  To understand how $\nu$
is related to the other critical exponents
finite-size scaling must be considered, since the role of the
correlation length, and in particular of its divergence, is strongly
related with another important quantity: the size $N$ of the network.

\subsection{Finite-size scaling}

The percolation phase transition is strictly defined only in infinite-size systems.

\change {The existence, in the thermodynamic limit, of a continuous phase transition
is manifested in finite systems} by fluctuations of the order parameter showing,
as a function of $\phi$, a maximum for a size-dependent value $\phi_c(N)$.
The amplitude of this maximum increases with the system size $N$.
At $\phi_c(N)$ the correlation length is
bounded by the maximum size that can be reached in the system, that is, in $D$ dimensions,
$\xi \sim N^{1/D}$.
Observing a finite system at $\phi_c(N)$ is equivalent to observing
an infinite system off-criticality at $\phi=\phi_c(N)$:
in both cases the correlation length is finite and $\xi \sim N^{1/D}$.
This defines $\phi_c(N)$ as an effective threshold
which tends to $\phi_c$ in the large-$N$ limit.
From Eq.~\eqref{xi} then, using $t=\phi_c(N)-\phi_c$,
\[N^{\frac{1}{D}} \sim |\phi_c(N)-\phi_c|^{-\nu } \implies |\phi_c(N)-\phi_c|\sim N^{-\frac{1}{D\nu} }. \]
Percolation critical properties are described by an effective field
theory, see Appendix~\ref{appendix:UCD}.
For such a theory the critical behavior in a space of dimensionality $D \geq D_{\text{UC}}$
larger than the {\em upper critical dimension}
is the same (mean-field) behavior observed at
$D=D_{\text{UC}}$~\footnote{Actually there are logarithmic corrections exactly at $D=D_{\text{UC}}$.}.
For standard percolation on lattices
the upper critical dimension is $D_{\text{UC}}=6$. Hence
the scaling of the effective threshold is in general
\begin{equation}
  |\phi_c(N)-\phi_c|\sim N^{-1/{\bar \nu} },
  \label{eq:definition_barnu}
\end{equation}
where ${\bar \nu} = \nu D$ for $D<D_{\text{UC}}$ while ${\bar \nu} =  \nu D_{\text{UC}}$ for
$D \ge D_{\text{UC}}$\change{, see \cite{dorogovtsev2022nature,privman1990finite}}.

\change{
This argument is also valid for random locally tree-like networks, where the dimension $D$ of the space in which a finite $N$ network can be embedded grows with $N$. For this reason, in the large $N$ limit, $D(N) > D_{\text{UC}}$, and the standard mean-field picture is expected to correctly capture the critical properties of the system.
}

At the effective threshold a GC starts to form
\begin{equation}
 \frac{\langle \mathcal{S} \rangle \big \rvert_{\phi_c(N)}}{\phi_c(N)}   \sim
N^{\theta}.
\label{eq:definition_theta}
\end{equation}
Close to criticality [$N \to \infty$, $\phi_c(N) \to \phi_c$], 
\begin{equation}
P^\infty = \frac{\langle \mathcal S \rangle\big \rvert_{\phi_c(N)}  }{\phi_c(N) N}
  \sim N^{\theta-1} 
\end{equation}
This quantity must also scale as $P^\infty \sim t^\beta$; since $t \sim N^{-1/{\bar \nu}}$ this 
implies
\begin{equation}
  \theta= 1-\frac{\beta }{\bar{\nu}}.
  \label{theta}
\end{equation}
The exponent $\theta$ can be seen also as the ratio between the fractal dimension $D_f$ of the GC and the embedding dimension $D$.

Fluctuations instead scale as
\begin{equation}
  \frac{\langle (\delta \mathcal{S})^2 \rangle \big \rvert_{\phi_c(N)} }{\phi_c(N) N} \sim t^{-\gamma'} \sim N^{\gamma'/\bar{\nu}}.
  \label{fluc}
\end{equation}
For standard percolation the mean square fluctuations of the order
parameter have the same critical singularity as the mean finite
cluster size, i.e., $\gamma' = \gamma$, see, e.g.,
Ref. \cite{dorogovtsev2022nature}. (In general this need not be the
case, and in more complex percolation problems---in particular, ones
involving hybrid phase transitions---the two exponents may be
different \cite{lee2016hybrid, lee2016critical}.)

Another scaling relation can be found as follows.
Close to criticality the largest clusters of size $s_\xi$ are fractals,
$s_\xi \sim \xi^{D_f} \sim t^{-D_f\nu}$.
Since by definition $s_\xi \sim t^{-1/\sigma}$ it follows that
$D_f = \frac{1}{\sigma \nu}$
implying
\begin{equation}
  \theta = \frac{1}{\sigma \bar{\nu}}.
\label{thetahs}
\end{equation}

Equating the two expressions for $\theta$ above implies the
hyperscaling relation
\begin{equation}
  2 \beta +\gamma=\bar{\nu}.
  \label{eq:hyperscaling}
\end{equation}

These relations allow us to determine all critical exponents from
numerical simulations.  In particular, $\bar{\nu}$ and
$\gamma/\bar{\nu}$ can be obtained by measuring the position and the
height of the peak of $\langle (\delta \mathcal{S})^2 \rangle$, while
$\theta$ can be obtained from the scaling of $\langle \mathcal{S}
\rangle$ at $\phi_c(N)$.  Once $\theta$, $\bar{\nu}$ and $\gamma$ are
known, $\beta$ and $\sigma$ can be obtained from the scaling
relations, and all the other critical exponents from them.

The scenario just described does not strictly apply for percolation
on highly hetereogeneous power-law distributed networks.
In such a case, as it will be shown below, the hyperscaling relation
is violated.

\change{
\subsection{Critical amplitude ratios}
Apart from critical exponents, also amplitudes of critical
behaviors obey universal properties. It is possible
to define~\cite{aharony1980universal} several amplitude ratios whose
values mark the universality class of the transition.
In particular we will be interested in the quantities $C_+/C_-$,
$\Xi_+/\Xi_-$ and $R_\chi = C_+ E^{-\delta} B^{\delta-1}$.
}

\change{
\subsection{Summary of the definitions}
We summarize in Table~\ref{tab:summary} definitions and physical meaning of all the critical exponents introduced in this section.
}

\begin{widetext}

 \change{
 \begin{table}[h]
  \begin{center}
  \change{
    \caption{Summary of the definitions of the various critical exponents. In the first seven lines the thermodynamic critical exponents; all the scalings hold for $t\ll1$, $h\ll1$. In the last two lines, critical exponents related to finite-size scaling, valid for large $N$.}
    \label{tab:summary}
    \begin{tabular}{|c|c|c|} \hline
       Exponent & Definition & Physical meaning \\ \hline
      $\alpha$ & $F \simeq A |t|^{2-\alpha}$ & Mean number of finite clusters per active node, see Eq.~\eqref{F}.\\
      $\beta$ & $P^{\infty}/\phi \simeq B t^{\beta}$ & Relative size (per active node) of the GC, see Eq.~\eqref{eq:P_infty_def}.\\
      $\gamma$ & $\langle s \rangle -1 \simeq C_{\pm}|t|^{-\gamma}$ & Mean size of small components to which a randomly chosen active node belongs, see Eq.~\eqref{eq:average_s_def}.\\
      $\delta$ & $1-\Psi(\phi_c,h) \simeq E h^{1/\delta}$ & Singular behaviour of the $h$-derivative of $\Omega(t,h)$ at criticality, see Eq.~\eqref{eq:definition_delta}.\\
      $\tau$ & $n_s(t) \simeq q_0s^{-\tau}f(q_1s^{\sigma}t)$ & Scaling of the large $s$ tail of the cluster size distribution, see Eq.~\eqref{eq:scaling}\\
      $\sigma$ & $s_{\xi}\simeq \Sigma |t|^{-1/\sigma}$ & Correlation size for finite components, see Eq.~\eqref{eq:definition_correlation_size}\\
      $\nu $ & $\xi \simeq \Xi |t|^{-\nu}$ & Average distance between nodes in the same finite component, see Eq.~\eqref{xi}. \\ \hline
      $\bar{\nu}$ & $|\phi_c - \phi_c(N)| \sim N^{-1/\bar{\nu}}$ & Effective finite-size threshold $\phi_c(N)$, see Eq.~\eqref{eq:definition_theta}. \\
      $\theta$ & $\frac{\langle \mathcal{S} \rangle\rvert_{\tiny \phi_c(N)}}{\phi_c(N)}   \sim N^{\theta} $ & Average largest cluster at the effective threshold $\phi_c(N)$, see Eq.~\eqref{eq:definition_barnu} \\ \hline
    \end{tabular}
    }
  \end{center}
\end{table}
}
\end{widetext}

\section{Generating function approach for uncorrelated random graphs}
\label{sec:recipe}

In this Section we calculate all the critical exponents for
percolation on uncorrelated random graphs, by analyzing the singular
behavior of generating functions. This strategy was already developed
in part in Refs.~\cite{newman2001random,cohen2002percolation}.
The treatment here is for an ensemble of uncorrelated\change{ locally tree-like} networks with
generic degree distribution $p_k$ (excess degree distribution
$q_k=(k+1)p_{k+1}/\langle k \rangle$)
and completely random in all other respects.\change{ We denote by $g_0(z)=\sum_{k}p_k z^k$ and $g_1(z) = \sum_{r}q_r z^r$ the
generating functions of the network's degree and excess degree distributions,
respectively.}

The application of the formulas derived in this Section to the specific
case of homogeneous or heterogeneous random networks is presented in
Appendix~\ref{AppendixCalculations} and the results are summarized in
Sec.~\ref{results}.

The generating function of the distribution $\pi_s(\phi)$ is
\begin{equation}
  H_0(z) \equiv \sum_{s=1}^{\infty}\pi_s(\phi) z^s
\end{equation}
For uncorrelated \change{locally tree-like} random graphs, it is well
known~\cite{callaway2000network,newman2007component}
\footnote{\change{Eq.~\eqref{eq:gen_clusters}
  and~\eqref{eq:recursive_H} correspond to the
  equations in~\cite{callaway2000network} via the mapping
  $H_0^{\text{\cite{callaway2000network}}}(z)=1-\phi+\phi H_0(z)$,
  $H_1^{\text{\cite{callaway2000network}}}(z)=1-\phi+\phi H_1(z)$.
  This is because we conditioned $H_0(z)$ and $H_1(z)$ on picking active nodes
  (see Eq.~\eqref{ns}),
  while the authors in \cite{callaway2000network} did not:
  hence the factor $\phi$ and the additional term $\pi_0=\rho_0=1-\phi$
  expressing the probability of picking an inactive node.
}}
that $H_0(z)$ obeys \change{\begin{equation}
    H_0(z) = zg_0\big( \phi H_1(z) +1-\phi \big) \equiv
    zG_0\big(H_1(z) \big) ,
 \label{eq:gen_clusters}
\end{equation}
}
where\change{ we defined $G_0(x) \equiv g_0(\phi x + 1 - \phi)$, and}
\begin{equation}
  H_1(z) \equiv \sum_{s=1}^{\infty}\rho_s(\phi) z^s
\end{equation}
is the generating function of $\rho_s(\phi)$, the distribution of the
total number of nodes reachable via a randomly chosen edge \change{that leads to an active node}.  Note that
the giant component, if there is one, is excluded from $H_1(z)$, i.e.,
$u \equiv H_1(1)$ is the probability that a randomly chosen edge, \change{ending in an active node}, leads
to a \change{finite} component of any size.  $H_1(z)$ is determined by the
implicit equation
\change{
\begin{equation}
 H_1(z) = zg_1\big(\phi H_{1}(z) + 1-\phi \big) \equiv   zG_1\big(H_1(z) \big),
 \label{eq:recursive_H}
\end{equation}
where we defined $G_1(x) \equiv g_1 (\phi x + 1-\phi)$.
}

The percolation threshold is determined by the
condition $1=\phi_c g_1'(1)$ \change{\cite{callaway2000network}}, implying $\phi_c=b^{-1}=\langle k
\rangle/(\langle k^2 \rangle -\langle k \rangle)$ \change{\cite{molloy1995critical,molloy1998size}}.
This holds if the network has a finite branching factor $b$.
If the branching factor diverges in the thermodynamic limit, then only the
percolating phase exists for any value of $\phi$, i.e., $\phi_c=0$.

By setting $z=e^{-h}$, and $m=1-H_1(e^{-h})$
Eqs.~\eqref{eq:gen_clusters} and~\eqref{eq:recursive_H}
can be rewritten as
\begin{align}
 \label{eq:psi_singular_UCM}
 \Psi(\phi,h)= &~
 e^{-h}g_0\big(1-\phi m \big) ,\\
m = &~ 1 - e^{-h}g_1\big(1-\phi m \big).
\label{eq:psi_recursive}
\end{align}

\subsection{The exponents $\beta$ and $\delta$}

Solving Eq.~\eqref{eq:psi_recursive} for $m(\phi, h)$ and
inserting the solution into Eq.~\eqref{eq:psi_singular_UCM}
allows to determine the exponents $\beta$ and $\delta$.
Everything depends on the behavior of the generating functions $g_0(z)$
and $g_1(z)$ close to $z=1$.
For $x \ll 1$ it is always possible to write
\begin{align}
\label{g0}
  g_0(1-x) &\simeq 1-\langle k\rangle x +\frac{1}{2}\langle k \rangle b x^2 +\{g_0(1-x)\}_{\text{sing}},\\
 g_1(1-x) &\simeq 1- bx +\frac{1}{2}d x^2 +\{g_1(1-x)\}_{\text{sing}},
 \label{g1}
\end{align}
where $\langle k \rangle$ is the average degree, 
always assumed to be finite, and $b, d$ are constants. Their values are
$b=\langle k(k-1) \rangle/\langle k\rangle$,
$d=\langle k(k-1)(k-2) \rangle/\langle k\rangle$ if the degree distribution
has finite second
and third moment, respectively. Otherwise $b$ and $d$ are numerical values
depending on the degree distribution, and the divergence of the moments
shows up in the singular part of the generating function.

We will always consider networks with a finite average degree $\av{k}$. Hence the singular contribution in $g_0(1-x)$ can always be neglected
because the finiteness of $\langle k \rangle$ implies that
$\langle k \rangle x \gg \{g_0(1-x)\}_{\text{sing}}$
for $x \to 0$ [in other words, $g_0(z)$ is continuous and differentiable in $z=1$,
and any singularity can appear only in the second derivative
or higher].
We can then always consider only the constant and the linear terms in the
r.h.s. of Eq.~\eqref{eq:psi_singular_UCM} so that
\begin{equation}
 \Psi(\phi,h)\simeq 1 - \langle k \rangle \phi m(\phi,h) - h.
 \label{eq:Psi_general}
\end{equation}

\change{Expanding Eq.~\eqref{eq:psi_recursive} for $|t| \ll 1$ and $h \ll 1$, we can obtain an approximate solution for $m(\phi,h)$. Plugging such a solution} into Eq.~\eqref{eq:Psi_general} then allows us to evaluate the
exponents $\beta$ and $\delta$ and the amplitudes $B$ and $E$.

\subsection{The exponent $\alpha$}
From Eq.~\eqref{eq:N_t_integral}, using the expression for $\Psi(\phi,h)$
in Eq.~\eqref{eq:psi_singular_UCM}, yields,
after integrating by parts~\cite{newman2018networks},
\begin{equation}
 F(\phi)=\Psi(\phi,0)-\phi \int_{0}^{\infty}dh e^{-h} g_0'(1-\phi m)
 \frac{\partial m}{\partial h}.
\end{equation}
Then, $g_0'(x)=\langle k \rangle g_1(x)$, and
Eq.~\eqref{eq:psi_recursive}
\begin{equation}
 F(\phi)=\Psi(\phi,0)-\phi \int_{0}^{\infty}dh (1-m) \frac{\partial
   m}{\partial h}.
\end{equation}
After the change of variable $x = m(\phi,h)$ at fixed $\phi$, and the
simple integration, we get
\begin{equation}
 F(\phi)=\Psi(\phi,0)-\frac{\phi \langle k \rangle}{2}[1-m(\phi,0)]^2,
 \label{eq:N_t_random_graphs}
\end{equation}
where $m(\phi,0)$ is the solution of Eq.~\eqref{eq:psi_recursive} for
$h=0$\footnote{This quantity, $m(\phi,0)$, is the same used to
  evaluate $P^\infty(\phi)$ and $\beta$: it is the probability of
  reaching the GC following a randomly chosen edge \change{ending in an active node}.}
Once the behavior of $m(t,0)$ for small $t$ is known, 
expanding Eq.~\eqref{eq:N_t_random_graphs} for small $t$, 
allows to determine the exponent $\alpha$.

\subsection{The exponent $\gamma$}

We are interested in the average size of the small
cluster to which a randomly chosen \change{active} node belongs, given by Eq.~\eqref{s}.
The numerator of this quantity, $\chi(\phi,0)$, may diverge at criticality
because of the power-law tail in $n_s(\phi_c)$ and determine the exponent $\gamma$,
while $\Psi(\phi, h)$ is always finite.
From Eq.~\eqref{eq:psi_singular_UCM} and from the
relation $\chi(\phi,h)=-\partial \Psi(\phi,h)/\partial h$ we find
\begin{equation}
 \chi(\phi, h) = \Psi(\phi,h) + e^{-h}g_0'(1-\phi m) \phi\frac{\partial m}{\partial h},
 \label{eq:chi}
\end{equation}
where $m$ is the solution of Eq.~\eqref{eq:psi_recursive}, from which,
\[\frac{\partial m}{\partial h} = e^{-h}g_1(1-\phi m)+e^{-h}g_1'(1-\phi m) \phi \frac{\partial m}{\partial h} ,\]
whose solution is
\begin{equation*}
 \frac{\partial m}{\partial h} = \frac{e^{-h}g_1(1-\phi m
   )}{1-e^{-h}\phi g_1'(1-\phi m)}=\frac{1-m}{1-e^{-h}\phi g_1'(1-\phi
   m)}.
\end{equation*}
Using $g_0'(x)=\langle k \rangle g_1(x)$, and again Eq.~\eqref{eq:psi_recursive}, we finally
obtain, from Eq.~\eqref{s},
\begin{equation}
  \langle s \rangle-1 =  \frac{\phi}{\Psi(\phi,0)} \frac{\langle k \rangle
   (1-m)^2}{1-\phi g_1'(1-\phi m)}.
\label{sminus1}
\end{equation}
Inserting the expression of $m(\phi,0)$ into Eq.~\eqref{sminus1}
and expanding for small $t=\phi-\phi_c$ provides
the singular behavior as criticality is approached.

\subsection{The exponent $\nu$}

Using Eq.~\eqref{xi2nets},
the problem of deriving the correlation length $\xi$ is reduced
to the computation of $\mu(l)$.
If there is no GC, all nodes are in \change{finite} components and since the network is
uncorrelated and locally tree-like, we have
\begin{equation}
  \mu(l)=\langle k \rangle \phi^l b^{l-1}.
\label{mulnoGC}
\end{equation}
Hence, using the formula
$\sum_{l \geq 0} l z^{l-1} =(1-z)^{-2}$,
we get
\begin{equation}
\xi^2 = \frac{\phi \langle k \rangle [1-\phi b ]^{-2}}{\chi(\phi,0)}.
\end{equation}

Here we generalize this argument to the case in which a giant
component is present, so that in the computation of $\mu(l)$ one must
explicitly impose that the GC is not reached. Let us define $P_l(n)$
as the probability that in the shell at distance $l$ from a given
active node, there are exactly $n$ nodes, in the same \change{finite} component
to which the original active node belongs. Then
$\mu(l)=\sum_{n}nP_l(n)=L_l'(1)$, where $L_l(z)=\sum_{n}P_l(n)z^n$ are
the generating functions of the distributions $P_l(n)$. Writing down an
explicit expression for $P_l(n)$ is rather involved. However, it is possible to
write, proceding step by step, a recursive expression for the generating
functions $L_l(z)$.
First of all, at level $0$
$P_0(n)=\delta_{1,n} \Psi(\phi,0)$, since an active node at distance
$0$ from a given active node, (i.e., itself) is in a \change{finite} component
with probability $\Psi(\phi,0)$. At distance $1$ it is easy to recognize that
\begin{equation}
 P_1(n)=\sum_{k}p_k {k \choose n} (\phi u)^{n}(1-\phi)^{k-n},
 \label{eq:P_1_n}
\end{equation}
where we remind the reader that $u$ is the probability that following
a randomly chosen edge we reach a \change{finite} component.  Multiplying
Eq. (\ref{eq:P_1_n}) by $z^n$ and summing over $n$ we obtain
\begin{equation}
 L_1(z)=g_0(\phi u z + 1-\phi)=G_0(uz).
\end{equation}
Let us consider now the nodes at distance $2$. It is possible to condition
on the number of nodes which were reachable at the previous step by writing
\begin{equation}
 P_2(n)=\sum_{m}P_2(n|m)P_1(m).
 \label{eq:P_2_n}
\end{equation}
The expression for $P_2(n|m)$ is simply given by
\begin{widetext}
\begin{equation}
 P_2(n|m) = \frac{1}{u^m}\sum_{n_1,\dots, n_m}\delta_{n,n_1+\dots+n_m} \sum_{r_1,\dots,r_k}q_{r_1}\dots q_{r_m}{ r_1 \choose n_1}\dots {r_m \choose n_m} (\phi u)^{n_1}\dots(\phi u)^{n_m}(1-\phi)^{r_1-n_1+\dots+r_m-n_m}.
\end{equation}
\end{widetext}
Note the factor $u^m$ in the denominator. It serves to correctly count
the number of $u$'s in the expression. In $P_1(m)$ there is a factor
$u$ for each of the $m$ branches, and at the second level these factors $u$
are replaced with the probabilities coming from the
branches encountered in the further exploration of the
tree. Multiplying Eq.~\eqref{eq:P_2_n} by $z^n$ and using the expression
for $P_2(n)$, a straightforward computation gives
yields
\begin{equation}
 L_2(z)=L_1(G_1(uz)/u)=g_0\big(\phi g_1(\phi u z + 1-\phi)+1-\phi\big).
\end{equation}
This argument can be used at any shell $l$, by conditioning on the
previous shell $l-1$ to have $m$ active nodes in \change{finite} components. We
arrive at the recursive equations
\begin{align}
 L_l(z)&=L_{l-1}\big(G_1(uz)/u \big),\\
 L_1(z)&=G_0(uz).
\end{align}
Essentially, as we proceed, we replace all the factors $uz$ 
at level $l-1$ with $G_1(u z)$. Taking the derivative of these
recursive equations, and evaluating them at $z=1$, leads to, recalling
that $u=G_1(u)$,
\begin{align}
 \mu(l)&=\mu(l-1)\phi g_1'(\phi u + 1 - \phi),\\
 \mu(1)&=\phi \langle k \rangle u^2,
\end{align}
whose solution is simply given by
\begin{equation}
 \mu(l) = \phi \langle k \rangle u^2 [\phi g_1'(\phi u + 1 - \phi)]^{l-1}.
 \label{eq:mu_l}
\end{equation}
Note that this expression reduces to the one shown before if $t<0$,
that is for $u=1$, i.e., in the absence of a GC.
Eq.(\ref{eq:mu_l}) is formally analogous to Eq.~\eqref{mulnoGC}, but with an
``effective'' branching factor which is reduced by the presence of the GC.
Inserting Eq.~\eqref{eq:mu_l} into $\chi(\phi,0)=\sum_{l \geq 0}
\mu(l)$,
\begin{align}
\nonumber
 \chi(\phi,0)&=\Psi(\phi,0)+\phi \langle k\rangle u^2 \sum_{l \geq 1}\left[\phi g'_1(\phi u + 1-\phi) \right]^{l-1} \\
 &=\Psi(\phi,0)+\frac{\phi \langle k \rangle u^2}{1-\phi g'_1(\phi u + 1-\phi)},
\end{align}
which coincides with Eq.~\eqref{eq:chi} evaluated at $h=0$.
For $\xi^2$ we have, summing again the derivative of the geometric series,
\begin{equation}
  \xi^2 = \frac{\phi \langle k \rangle u^2 [1-\phi g'_1(\phi u + 1-\phi)]^{-2}}{\Psi(\phi,0)+\phi \langle k \rangle u^2[1-\phi g'_1(\phi u + 1-\phi)]^{-1}}.
  \label{xi2}
\end{equation}
Expanding the numerator and the denominator for $t$ close to $0$, both
below ($u=1$) and above ($u<1$) the transition, allows us to compute
the critical exponent $\nu$.

\subsection{The exponents $\sigma$ and $\tau$}

From the singular behavior of $H_0(z)$ and $H_1(z)$, also
the scaling form of $\pi_s(\phi)$ and hence the exponents $\tau$ and $\sigma$
can be worked out explicitly.
Indeed, if the generating function $H_0(z)$ has a convergence radius
$\rho_{H_0}$,
\begin{equation}
  H_0(z) = A_0-B_0(\rho_{H_0}-z)^{a_0},
  \label{H0sing}
\end{equation}
inversion (Tauberian) theorems~\cite{Flajolet2009} guarantee that,
for $s \gg 1$, $\pi_s$ scales as
\begin{equation}
  \pi_s(\phi) \simeq \frac{B_0}{\Gamma(-a_0)} s^{-a_0-1} e^{-s/s_\xi},
  \label{pi_tauberian}
\end{equation}
in agreement with Eq.~\eqref{eq:scaling}, with $\tau=2+a_0$, and
\begin{equation}
  s_{\xi}=\frac{1}{\log(\rho_{H_0})}.
  \label{eq:definition_s_xi}
\end{equation}
Hence in order to find $\sigma$ and $\tau$ it is necessary to
determine the singular behavior of $H_0(z)$ which,
from Eq.~\eqref{eq:gen_clusters}, is related to the singular
behavior of $H_1(z)$, that we now analyze.

By setting $u=H_1(z)$, Eq.~\eqref{eq:recursive_H} can be rewritten as
\begin{equation}
z=\psi(H_1(z)),
\label{eq:psi}
\end{equation}
where $\psi(u)=u/G_1(u)$ is the inverse function of $H_1(z)$.
Assuming $G_1(0)\neq  0$ (as it is always the case for $0 < \phi \le 1$)
and $\lim_{u \to R^-} uG_1'(u)/G_1(u) >1$
[where $R$ is the convergence radius of $G_1(u)$]
there exists (see Ref.~\cite{Flajolet2009}, Proposition IV.5, page 278)
a unique solution $u^* \in (0,R)$
of the  \textit{characteristic equation} $\psi'(u^*)=0$, that is, of the equation
\begin{equation}
 G_1(u^*)-u^*G_1'(u^*)=0.
 \label{eq:characteristic}
\end{equation}

Then, the convergence radius of the power series $H_1(z)$ is
\begin{equation}
\rho_{H_1} = \psi(u^*) = \frac{u^*}{G_1(u^*)}.
\end{equation}

In other words, the inversion of the function $\psi(u)$ can be performed
only up to $u^*$, where its first derivative vanishes.
As a consequence, the function $H_1(z)$ has a singularity at
$z=\rho_{H_1}=\psi(u^*)$, and this is the closest singularity to the origin.

The singular behavior of $H_1(z)$ close to $\rho_{H_1}$
is
\begin{equation}
  H_1(z) \simeq u^*-A_1(\rho_{H_1}-z)^{a_1}, ~~~z \to \rho_{H_1}^-,
  \label{H1sing}
\end{equation}
with $A_1$ a positive constant\footnote{There are logarithmic corrections to Eq. (\ref{H1sing}) if $a_1$ is an integer.}.
Note that $H_1(z=\rho_{H_1}) = H_1(\psi(u^*))=u^*$.

If $G_0(H_1)$ is analytic at $H_1 = u^*$, insert Eq.~\eqref{H1sing}
into Eq.~\eqref{eq:gen_clusters} and expanding $G_0(H_1(z))$ leads to
\begin{equation}
 H_0(z)  \simeq \rho_{H_1}G_0(u^*)-\rho_{H_1}G_0'(u^*)A_1(\rho_{H_1}-z)^{a_1},
 \label{eq:leading_singularity_H_0}
\end{equation}
that is, Eq.~\eqref{H0sing} with $\rho_{H_0}=\rho_{H_1}$, $A_0=\rho_{H_1} G_0(u^*)$,
$B_0 = \rho_{H_1} G_0'(u^*) A_1$ and $a_0=a_1$.

The case in which $G_0(H_1)$ is nonanalytic at $H_1= u^*$
may occur when $u^*=1=\rho_{H_1}$.
In such a case we can insert Eq.~\eqref{H1sing}
into Eq.~\eqref{eq:gen_clusters} and use now the singular expansion
for $G_0(H_1)$ around $u^*=1$, see Eq.~\eqref{g0}. Since the
singularity in $G_0(H_1)$ has an exponent larger than 1,
the leading singularity in $H_0(z)$ is still given by 
Eq.~\eqref{eq:leading_singularity_H_0}.

From now on we denote $\rho_{H_0}=\rho_{H_1}=\rho$.
$H_1(z)$ is the generating function of a probability distribution, hence its
convergence radius must be $\rho \ge 1$
(the coefficients cannot diverge exponentially).

Recalling that $u=H_1(z)$, the singular behavior of $H_1(z)$
[Eq.~\eqref{H1sing}] is found by determining how $H_1(z)-u^*=u-u^*$
depends on $\rho-z$.
Since $\rho=\psi(u^*)$ and $z=\psi(u)$ this implies that we
have to write
\begin{equation}
  \rho-z = \psi(u^*)-\psi(u),
\label{inversion}
\end{equation}
expand the r.h.s. for small $\epsilon=u^*-u$ and invert
to obtain how $\epsilon$ is singular as a function of $\rho-z$:
$\epsilon \sim (\rho-z)^{a_1}$. The exponent $a_1=a_0$ is the one
appearing in Eq.~\eqref{H1sing} and hence it determines $\tau$.
The type of singularity obtained by this inversion depends on
the expansion of the function $\psi(u)$, which in turn strongly
depends on the degree distribution and whether 
supercritical ($t>0$), critical ($t=0$) or subcritical ($t<0$)
properties are considered, as shown in detail in Appendix~\ref{appendix:psi_asymptotic}.

\subsection{The exponent $\bar \nu$}

Consider a network with finite maximum degree $k_c$ and
large $N$.
Regardless of the degree-distribution, such a system is effectively
homogeneous. \change{In the infinite-size limit, this system has a critical point $\phi_c(k_c)=b(k_c)^{-1}$ which depends on the maximum degree $k_c$.}
In \change{the denominator of} Eq.~\eqref{xi2}, \change{for $\phi$ close to $\phi_c(k_c)$},
$\Psi(\phi,0)$ can be neglected, leading to (see
Appendix~\ref{AppendixCalculations})
$\xi^2 \simeq |1-\phi b(k_c)|$.
Hence
\begin{equation}
\xi \simeq b(k_c)^{-1/2} |\phi-b(k_c)^{-1}|^{-1/2}.
\end{equation}

\change{According to our finite-size scaling argument, t}he effective critical point is determined by this expression, evaluated
at $\phi=\phi_c(N)$, with the left hand side proportional to $N^{1/6}$, since the
system is homogeneous \change{and $D_{\text{UC}}=6$}. Neglecting for simplicity the absolute value,
this yields
\begin{equation}
  \phi_c(N,k_c) \simeq b(k_c)^{-1} \left(1+ C_1 N^{-1/3}\right),
  \label{phic0}
\end{equation}
where $C_1$ is a constant. \change{If $k_c$ does not depend on $N$, or the dependence is subalgebraic, Eq.~\eqref{phic0} leads to $\bar{\nu}=3$. However, if $k_c \sim N^{\omega}$, as for heterogeneous networks, a deviation from the standard mean-field exponent may occur.}

\section{Results}
\label{results}

The application of this general approach to the specific cases of
homogeneous and power-law degree distributions ($p_k \sim k^{-\gamma_d}$)
is detailed in Appendix~\ref{AppendixCalculations}. For what concerns thermodynamic critical exponents and amplitude ratios, results are summarized in
Table~\ref{tab:table1}. Note that all exponents are derived
independently, scaling relations are not used (yet they are verified
by our results except for the relation $\delta=1/(\tau-2)$ for $2<\gamma_d<3$).
Some of the values in Table~\ref{tab:table1}
\change{(for exponents $\tau$ and $\gamma$ in the case $2 <\gamma_d <3$)}
disagree with \change{some} values reported previously in the
literature. In the concluding section we discuss in detail the origin
of these discrepancies and what led to the previous incorrect claims.

\begin{widetext}

  \begin{table}[h]
  \begin{center}
    \caption{Values of the thermodynamic critical exponents and
      critical amplitude ratios for uncorrelated power-law distributed
      networks in the various $\gamma_d$ ranges, computed from the
      exact solution with the generating functions.~\change{
        Values indicated by $^a$ were first calculated in Ref.~\cite{cohen2002percolation}. Values indicated by $^b$ were first calculated in Ref.~\cite{Cohen2004}.
        Values indicated by $^c$ were first calculated in Ref.~\cite{newman2001random}.
        See Appendix~\ref{RchiPL} for a discussion of values indicated by $^*$.
    } }
    \label{tab:table1}
    \begin{tabular}{|c|c|c|c|c|c|c|c|c|c|c|} \hline
      Network type & $\beta$ & $\delta$ & $\alpha$ & $\gamma$ & $\nu$ & $\sigma$ & $\tau$& $C_+/C_-$& $\Xi_+/\Xi_-$& $R_{\chi}$  \\ \hline
      Homogeneous (and $\gamma_d>4$) & $1^a$ & $2$ & $-1$ & $1^a$ & $\frac{1}{2}^b$ & $\frac{1}{2}^a$ & $\frac{5}{2}^c$ & $1$ & $1$ & $1$\\
      $3<\gamma_d<4$ &$\frac{1}{(\gamma_d-3)}^a$ & $\gamma_d-2$ &$-\frac{5-\gamma_d}{\gamma_d-3}$ & $1^a$& $\frac{1}{2}^b$ & $\frac{\gamma_d-3}{\gamma_d-2}^a$ & $2+\frac{1}{\gamma_d-2}^a$ & $\frac{1}{\gamma_d-3}$ & $1$ & $\frac{1}{\gamma_d-3}$\\
      $2<\gamma_d<3$ & $\frac{1}{(3-\gamma_d)}^a$ & $1^*$ & $-\frac{3(\gamma_d-7/3)}{3-\gamma_d}$ & $-1^a$ & $ -\frac{1}{2} $& $\frac{3-\gamma_d}{\gamma_d-2}^a$ & $2+\frac{1}{\gamma_d-2}^a$ & N/A& N/A& $\frac{1}{3-\gamma_d}^*$ \\\hline
    \end{tabular}
  \end{center}
\end{table}

\end{widetext}

Beyond the value of the exponents, the analysis yields a very rich
picture for the cluster size distribution $n_s(t)$, with nontrivial
preasymptotic behaviors and violations of the scaling ansatz, as
summarized here.

\begin{figure}
    \includegraphics[width=0.98\columnwidth]{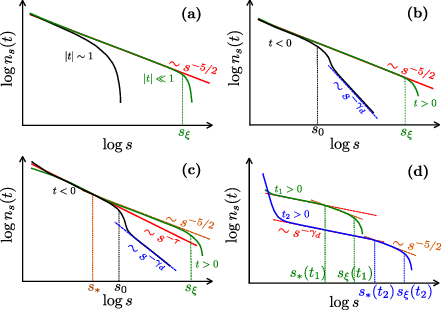}
    \caption{Figure with schematic log-log plots of
      $n_s(t)$ in the various regimes.
      (a) Homogeneous networks. The red line corresponds to
      the critical scaling $s^{-\tau}$ with $\tau=5/2$. For $|t| \ll
      1$, both positive and negative, an exponential cutoff occurs for
      for $s=s_{\xi}$ with the correlation scale $s_\xi$ diverging as
      criticality is approached.  (b) $\gamma_d>4$. The red line corresponds to
      the critical scaling $s^{-\tau}$ with $\tau=5/2$. For positive
      $t\ll 1$ (green line), the same picture as in (a) holds. However,
      for $|t|\ll 1$ but $t<0$ (black line), the asymptotic scaling is
      given by $s^{-\gamma_d}$ (blue dashed line), and the exponent
      $\tau$ is visible only up to the crossover scale $s_0$.  (c)
      $3<\gamma_d<4$. The red line corresponds to the critical scaling
      $s^{-\tau}$ with $\tau=2+1/(\gamma_d-2)$. For small positive $t$
      (green line), the asymptotic scaling is with
      an exponent $5/2$ (orange dashed line) followed by an
      exponential cutoff occurring at $s_{\xi}$, while in the
      preasymptotic regime $1 \ll s \ll s_{*}$ the scaling with
      exponent $\tau$ is observed. For small $t<0$ (black line) the
      same picture as in (b) holds.  (d) $2<\gamma_d<3$. Only the case
      $t>0$ is possible. The two cases $t_2<t_1 \ll 1$ exhibit the same
      scaling behavior: a preasymptotic scaling with exponent
      $\gamma_d$, followed (after the scale $s_*$) by the exponent
      $5/2$ and then an exponential cutoff at $s_{\xi}$.
      However, as $t$ is decreased, the amplitude of
      these decays decreases and vanishes asymptotically, leaving only
      $n_s(t=0)=\delta_{s,1}$.
    }
    \label{Figsummary}
\end{figure}

On homogeneous networks percolation is perfectly described by the
scaling ansatz~\eqref{eq:scaling} [see Fig.~\ref{Figsummary}(a)]: the
cluster size distribution is a power-law with exponent $\tau=5/2$
cut off exponentially at the correlation size $s_\xi$. Both for
positive and negative $t$, $s_\xi$ diverges as
$|t|^{-1/\sigma}$ as criticality is approached, with the mean-field
exponent $\sigma=1/2$.

The same scenario applies for power-law networks with $\gamma_d>4$ and $t>0$,
but for $t<0$ the picture is different [see Fig.~\ref{Figsummary}(b)]:
after an exponential cutoff, occurring for
$s_0  \sim |t|^{-1/\sigma}$,
the cluster size distribution decays indefinitely as $s^{-\gamma_d}$, a legacy of the degree distribution of the original network~\cite{newman2007component}.
This violation of the scaling ansatz Eq.~\eqref{eq:scaling}
is a direct effect of the heterogeneity of the substrate but it is
not related to different critical properties of the percolation
process. Indeed, for $t \to 0^{-}$ the asymptotic tail with exponent
$\gamma_d>\tau$ is not responsible for the divergence of the
fluctuations, which are determined instead by the preasymptotic scaling with
exponent $\tau$. The scale $s_0\sim |t|^{-1/\sigma}$ plays a role similar to
the correlation size $s_{\xi}$ for $t>0$ and scales in the same way.

For more heterogeneous networks, $3 < \gamma_d <4$, both $\sigma$ and
$\tau$ become $\gamma_d$-dependent, but the most relevant change is
that for $t>0$ the asymptotic decay of $n_s(t)$ is of the form
$s^{-5/2} e^{-s/s_\xi}$, i.e., it is different from the decay
$s^{-\tau}$ occurring exactly at criticality ($t=0$).  The way these
two apparently incompatible behaviors are reconciled is depicted in
Fig.~\ref{Figsummary}(c): a crossover scale $s_*$, smaller than
$s_\xi$ but diverging with $t$ in the same way, separates the
preasymptotic decay $s^{-\tau}$ from the regime $s^{-5/2} e^{-s/s_\xi}$.
As criticality is approached, both $s_*$ and $s_\xi$
diverge and the regime $s^{-\tau}$ extends to all scales.  For $t<0$
instead the phenomenology is analogous to the case for $\gamma_d>4$
only with a different value of $\tau$.

Finally, for $2<\gamma_d<3$ only the percolating phase $t>0$
  exists, as $\phi_c=0$.  In contrast to what happens for the other
  types of networks discussed above, when $\phi \to 0$ the system
  exhibits no critical behavior: $\langle s \rangle $ does not
  diverge ($\gamma=-1$), the cluster size distribution tends to be
  delta distributed ($n_s(0)=\delta_{s,1}$),
  and the correlation length $\xi$ goes to zero ($\nu = -1/2$).
  However, for small $t$ the cluster size distribution
  exhibits a nontrivial behavior, displaying
  some features of a critical transition, see Fig.~\ref{Figsummary}(d).
  For fixed small $t$, similarly to the case $3<\gamma_d<4$,
  $n_s$ exhibits two power-law decays, separated by a crossover scale $s_*$.
  For $1 \ll s \ll s_*$, it decays
  as $s^{-\gamma_d}$, while for $s \gg s_{*}$ we find $n_s \sim s^{-5/2}e^{-s/s_{\xi}}$,
  where $s_{\xi}\sim t^{-1/\sigma}$, with
  $\sigma=(3-\gamma_d)/(\gamma_d-2)$. Also the crossover scale $s_*$ diverges
  as $t^{-1/\sigma}$.
  Hence one could be tempted to conclude that $\tau=\gamma_d$,
  in agreement with some theoretical results in the
  literature~\cite{lee2004evolution},
  and in contrast with others~\cite{cohen2002percolation,li2021percolation}.
  Note that with this value the scaling
  relations~\eqref{scalingrels} would not be satisfied. 
  But crucially the large-$s$ tail of $n_s$ is multiplied
  by a $t$-dependent factor, vanishing as $t \to 0$,
  in agreement with the fact that $n_s(\phi) \to \delta_{s,1}$.
  Despite the fact that $\phi_c=0$ is not a veritable critical point
  it is still possible to write this behavior of $n_s(t)$ in a
  scaling form~\eqref{eq:scaling}, where now $\tau=2+1/(\gamma_d-2)$
  (in agreement with~\cite{cohen2002percolation} and with the scaling relations)
  and the scaling function $f_+(x)$ does not
  go to a constant for $x \to 0$, vanishing instead as $x^{\gamma_d-1}$.
  This nontrivial form of the scaling function $f_+(x)$ has the
  consequence that for fixed $t$ the cluster size distribution decays
  with an exponent $\gamma_d$ different from $\tau$ and a $t$-dependent
  prefactor vanishing as $t \to 0$.

  All these results about $n_s(t)$ are in agreement with what one can deduce
from the theory for the network component size distribution formulated
by Kryven~\cite{kryven2017general}, adapting the results therein by
assuming that nodes are diluted with probability $\phi$ (see
Appendix~\ref{Kryven}).

\begin{table}[h!]
  \begin{center}
    \caption{Values of size-related critical exponents for
      uncorrelated power-law distributed networks in the various
      $\gamma_d$ ranges, where the hard structural cutoff of the network is
      taken as $\kmax \sim N^{1/\omega}$, with $\omega$ a positive
      parameter obeying the constraint $\omega \geq 1$ ($\omega \geq 2$ for $2 <\gamma_d<3$).
      The homogeneous case corresponds to a
      sub-algebraically growing $k_c(N)$.
      The hyperscaling relation, Eq.~\eqref{eq:hyperscaling}, is satisfied
      when the values in the last two columns coincide (see Appendix~\ref{appendix:bar_nu}). ~\change{
        Values indicated by $^a$ were first calculated in Ref.~\cite{cohen2002percolation}.}}
      \label{tab:table2}
    \begin{tabular}{|cc|c|c|c|} \hline
      \multicolumn{2}{|c|}{Network type} & $\bar{\nu}$ & $\theta$ & $\frac{1}{\sigma\bar{\nu}}$\\ \hline
      Homogeneous & & $3^{a}$ & $\frac{2}{3}$  & $\frac{2}{3}^a$ \\
      $4<\gamma_d$ & $\omega \le 3(\gamma_d-3)$  & $3$ & $\frac{2}{3}$ & $\frac{2}{3}$ \\
       & $\omega \ge 3(\gamma_d-3)$ & $\frac{\omega}{\gamma_d-3}$ & $1-\frac{\gamma_d-3}{\omega} $  & $\frac{2(\gamma_d-3)}{\omega}$\\
      $3<\gamma_d <4$ & $\omega \le \gamma_d-1$ & $\frac{\gamma_d-1}{\gamma_d-3}$ & $\frac{\gamma_d-2}{\gamma_d-1}$ & $\frac{\gamma_d-2}{\gamma_d-1}$\\
       & $\omega \ge \gamma_d-1$ & $\frac{\omega}{\gamma_d-3}$ & $1-\frac{1}{\omega}$ & $\frac{\gamma_d-2}{\omega}$\\
      $2<\gamma_d<3$ & & $\frac{\omega}{3-\gamma_d}$ & $1-\frac{1}{\omega}$ &  $\frac{\gamma_d-2}{\omega}$\\\hline
    \end{tabular}
  \end{center}
\end{table}

Also concerning finite-size scaling the picture is surprisingly rich (see
Table~\ref{tab:table2}
 for a summary of results and Appendix~\ref{appendix:bar_nu} for a detailed discussion).
Contrary to naive expectation and to what is reported in the
literature~\cite{cohen2002percolation, dorogovtsev2008critical, li2021percolation}, the exponent $\bar \nu$, governing how the effective threshold approaches
the asymptotic limit, does not necessarily assume the mean-field value
$\bar \nu=3$ for power-law degree-distributed networks with $\gamma_d>4$.
Indeed, $\bar \nu$ assumes its mean-field value 3 only for
small enough $\omega$, the exponent governing how the network hard
structural cutoff $\kmax$ grows with the system size $N$, $\kmax \sim N^{1/\omega}$.
Specifically, for $\gamma_d > 4$ we have $\bar\nu = 3$ only for $\omega \le 3(\gamma_d-3)$.
For $\omega \ge 3(\gamma_d-3)$ the exponents $\bar \nu$ and
$\theta$ depend also on $\omega$ and hyperscaling relations do not hold.
The crossover value for $3 < \gamma_d < 4$, above which hyperscaling relations are violated, is $\omega = \gamma_d-1$.
For $2<\gamma_d<3$ the transition occurring at $\phi=0$ is not really
critical and the whole finite size scaling framework must be
interpreted differently (see Appendix~\ref{appendix:bar_nu}).

\section{Numerical simulations}
\label{simul}

We performed numerical simulations to test some of the analytical results,
in particular with regard to finite size scaling.
We considered networks built according to the uncorrelated configuration
model~\cite{catanzaro2005generation}, with $\kmax = \km N^{1/\omega}$,
$\omega=3$, $\km=3$ and various $\gamma_d$. Site percolation was simulated
by means of the efficient Newman-Ziff
algorithm~\cite{newman2000efficient} where for each network $M=1000$
realizations were run. To identify the critical point we considered two different
susceptibilities~\cite{castellano2016on}: $\phi_{c1}$ is the position
of the peak of $\chi_1 = (\av{{\cal S}^2}-\av{{\cal S}}^2)/(\phi N)$,
while $\phi_{c2}$ is the position of the peak of
$\chi_2 = (\av{{\cal S}^2}-\av{{\cal S}}^2)/\av{{\cal S}}$.
We evaluate the scaling of the size of the largest cluster at criticality
for both determinations of the critical point,
$S_1(N) = {\av{\cal S}}|_{\phi_{c1}}(N)/\phi_{c1}(N) \sim N^{\theta}$ and
analogously for $S_2(N)$.
The height of the peak of $\chi_1$ scales as $N^{\gamma/{\bar \nu}}$
[see Eq.~\eqref{fluc}], while the peak of $\chi_2$ diverges with
an exponent $(\gamma+\beta)/{\bar \nu}=1/(\sigma {\bar \nu})$.
This latter exponent coincides with $\theta$ if hyperscaling holds. We then average all the measured quantities over $100$ to $1000$ different realizations of the
network substrate.
We finally evaluate the effective exponent $\lambda$ of the observable $O \sim N^{\lambda}$ as a
function of size by performing simulations for several sizes $N_i$ and calculating
\begin{equation}
 \lambda(N_i) = \frac{\log({O}(N_{i+1})/{O}(N_i))}{\log(N_{i+1}/N_i)}.
 \label{eq:EffExp}
\end{equation}

\begin{figure}
    \includegraphics[width=0.97\columnwidth]{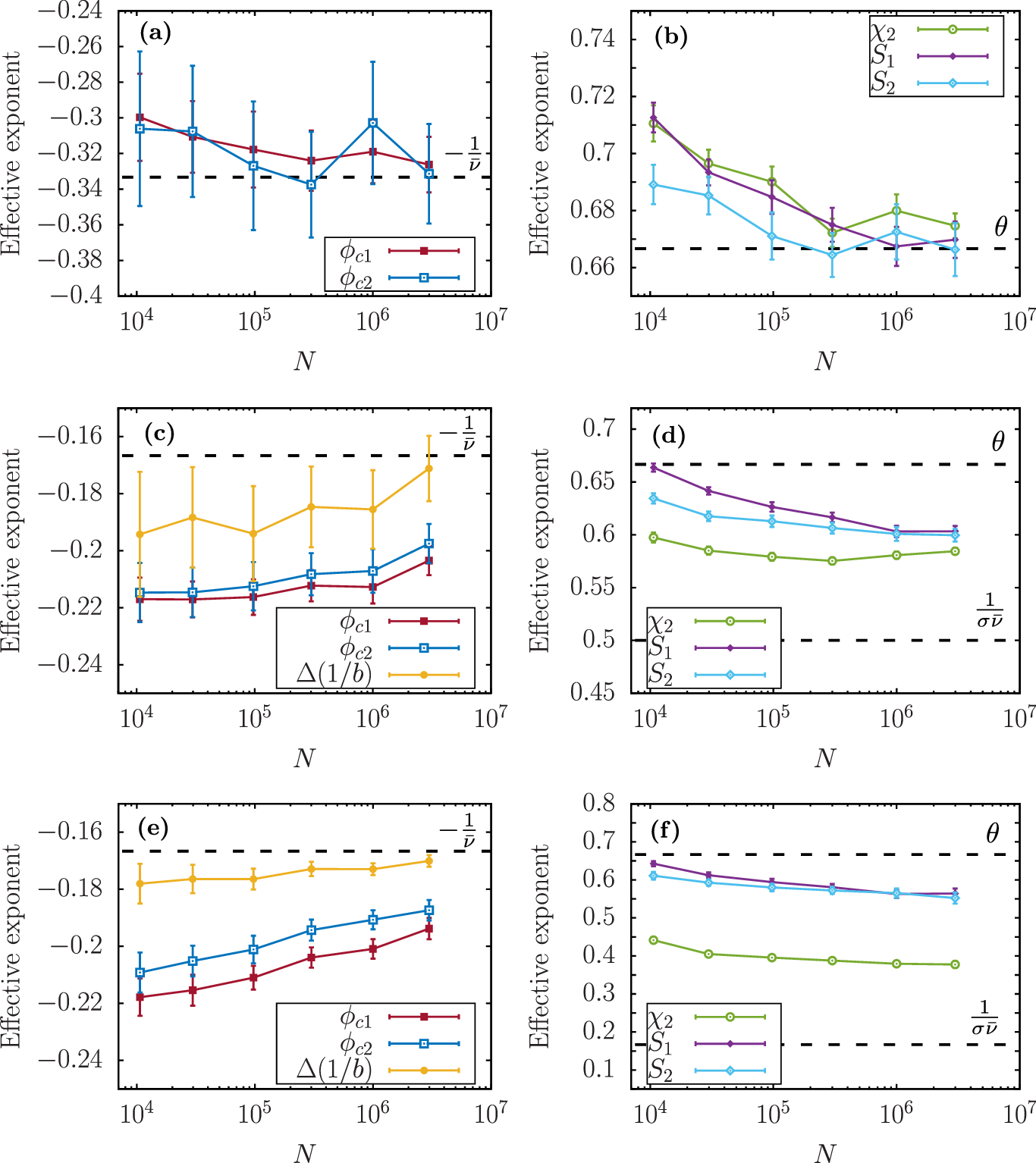}
    \caption{Effective exponents of the observables $\phi_{c1}$,
      $\phi_{c2}$, $\chi_2$, $S_1$, and $S_2$ measured via
      Eq.~\eqref{eq:EffExp} in numerical simulations for power-law
      networks, with (a),(b) $\gamma_{d}=5$, (c),(d) $\gamma_{d}=3.5$,
      and (e),(f) $\gamma_{d}=2.5$. Dashed lines correspond to the
      theoretical predictions of the exponent as in
      Table~\ref{tab:table2}. Note that in (b), where the hyperscaling
      is expected to hold, $\theta=1/(\sigma \bar{\nu})$. In (c) and
      (e), we also report the scaling of the branching factor measured
      over the network samples, showing the presence of preasymptotic
      effects.}
    \label{Figsimulations}
\end{figure}
For $\gamma_d=5$, finite size scaling is the same valid for homogeneous networks:
$\bar \nu=3$, $\theta=2/3$, and hyperscaling holds.
Fig.~\ref{Figsimulations}(a) confirms that this picture holds for networks
with size of the order of $N=10^6$ or larger.
For $\gamma_d=2.5$ instead, we do not expect hyperscaling to hold and the predicted
value of $\bar \nu$ is 6. While the violation of hyperscaling is clear in
Fig.~\ref{Figsimulations}(f), it is evident that the effective exponents are far
from the predicted asymptotic values. Effective exponents differ from those
predicted analytically also for $\gamma_d=3.5$ [Fig.~\ref{Figsimulations}(d)].
Also in this case strong preasymptotic effects dominate, as shown
by Fig.~\ref{Figsimulations}(c): while the scaling of $\Delta (1/b)=1/b(N)-1/b_\infty$
approaches the expected behavior $N^{-1/6}$ for the largest sizes considered,
the effective exponents derived from the scaling of $\phi_{c1}(N)$ and $\phi_{c2}(N)$
are close to $-0.2$. This indicates that the term proportional to $N^{-1/3}$
in Eq.~\eqref{phic} cannot be safely neglected, even for the largest values of $N$
considered. The effective exponent close to $-0.2$
can be interpreted, since $1/{\bar \nu} = (\gamma_d-3)/\omega$, as the
consequence of an effective value of $\omega \approx 2.5$ which, inserted into
$\theta=1-1/\omega$, accounts for the effective exponent of $S_1$ and $S_2$ close
to $0.6$.
The true asymptotic behavior could be observed only for system sizes such that
the blue and red curves in Fig.~\ref{Figsimulations}(c),(e) reach
the dashed horizontal line.
Networks of several orders of magnitude
larger than those we can consider would be needed.
For the same reason, it is impossible to verify numerically the breakdown
of hyperscaling and the validity of the prediction $\bar \nu = \omega/(\gamma_d-3)$ 
for $\gamma_d>4$ and $\omega \ge 3(\gamma_d-3)$.

\section{Conclusions}
\label{sec:conclusions}

In this paper we have provided a complete and coherent analysis of
the critical behavior of standard site and bond percolation models
on uncorrelated \change{locally tree-like} random networks with generic degree distribution $p_k$.
\change{Even if recent exact results about anomalous exponents on finite-$N$ scale-free networks \cite{dhara2018critical,dhara2021critical,bhamidi2021multiscale} suggest that a full mathematical theory is still lacking, percolation on random graphs} has been the subject of intense research activity for
decades and a general understanding of many nontrivial features
had already been reached a long time ago. In particular the vanishing threshold
for strongly heterogeneous networks and the dependence of some critical
exponents on the exponent $\gamma_d$ have been recognized over 20 years
ago~\cite{cohen2000resilience,cohen2002percolation}.
However, results about critical properties are scattered in a large body
of literature and
often interspersed with imprecise or incorrect statements.

Moreover a complete theory for all exponents and the scaling functions
was lacking.
In this work we have filled these gaps.
\change{Exploiting the locally tree-like nature of the networks considered, }by means of the generating function\change{s} approach we have derived all critical
properties for both infinite and finite systems.

In this way we have clarified the true value of some exponents
\change{for scale-free networks}, over
which confusion existed in the literature.  \change{In particular:
  the Fisher exponent $\tau$,
  that was claimed to be equal to
  $3$~\cite{dorogovtsev2008critical}, to
  $2+1/(\gamma_d-2)$~\cite{cohen2002percolation}, to
  $\gamma_d$~\cite{lee2004evolution} or even to be not
  defined~\cite{radicchi2015breaking} is found to be $2+1/(\gamma_d-2)$.
  The exponent $\gamma$, claimed to be not
  defined~\cite{radicchi2015breaking} turns out to be equal to $-1$.
  The exponent $\bar \nu$, governing finite size scaling, has an anomalous
  behavior (see Table~\ref{tab:table2}) at odds with the previously claimed values of
  $2/(3-\gamma_d)$~\cite{radicchi2015breaking} and
  $(\gamma_d-1)/(3-\gamma_d)$~\cite{cohen2002percolation}.}
We have derived the value of
other exponents \change{$(\alpha, \delta)$} not explicity calculated
before, and we have
determined for the first time critical amplitude ratios.

A crucial finding is the detailed understanding of how the cluster
size distribution $n_s(t)$ behaves in the various ranges of $\gamma_d$
values.  It turns out that the usual scaling assumption
$n_s(t) \sim s^{-\tau} f(s/s_\xi(t))$
with $f(x)=\exp(-x^{1/\sigma})$ is never fully correct for power-law distributed networks.  For $\gamma_d>4$ in the subcritical case the
exponential cutoff is followed by an asymptotic decay
$s^{-\gamma_d}$. For $3<\gamma_d<4$ this subcritical feature is
accompanied, above the threshold, by a crossover such that the
asymptotic decay occurs with an exponent that differs from the
critical value $\tau$.  For strongly heterogeneous networks
($2<\gamma_d<3$) the vanishing threshold implies that $\phi_c=0$ is
not a true critical point.  It is still possible to write $n_s(t)$ in
a scaling form but $f(x)$ vanishes for $x \to 0$.  One of the
consequences of this nontrivial scaling is that the decay of $n_s(t)$
for fixed $t$ is governed by an exponent $\gamma_d$ differing from the
Fisher exponent $\tau=2+1/(\gamma_d-2)$ and is multiplied by a factor
which vanishes as $t \to 0$. The claims in the literature that $\tau$
is equal to $\gamma_d$ or not even well defined reflect a partial
understanding of the scaling properties, that we have fully
elucidated here.

We have also presented a consistent theory for finite-size scaling
properties, showing that the exponent $\bar \nu$ may depend on how the
maximum degree $k_c$ diverges with the system size.  At odds with all
previous literature, we find that this may happen even for
$\gamma_d>4$, (when all thermodynamic exponents are equal to those for
homogeneous systems) provided $k_c$ diverges sufficiently slowly.
This dependence on the network maximum degree implies that
hyperscaling relations, usually assumed to be valid, are violated in
this case.

Our work may stimulate further in depth investigation
(or reinvestigation) of critical
properties for other types of percolation processes on
networks. What happens when nodes are removed
in a non fully random way, for example targeting first
highly~\cite{cohen2001breakdown} or
poorly~\cite{gallos2005stability,caligiuri2020degree} connected nodes?
Or if connected components are defined allowing for gaps in paths
connecting nodes (extended-range percolation~\cite{cirigliano2023extended,cirigliano2024general})? What can be said about critical exponents and scaling relations when the system undergoes a discontinuous or hybrid transition?

Our results may have implications for epidemic models and spreading processes in general, due to well-established links between such models and percolation theory \cite{pastor2015epidemic}.
Our generating functions approach may also be useful in the analysis of critical properties in other models of interacting systems on networks exhibiting continuous phase transitions, e.g., spin systems, models of synchronization and opinions dynamics \cite{dorogovtsev2008critical}.

\change{Finally, an interesting question is to understand how our results, obtained within the assumption of locally tree-like networks, are affected by the presence of clustering and short loops.}

The exploration and understanding of universal behaviors in models defined on complex networks -- often qualitatively different from those observed on regular topologies -- offer a pathway to deeper theoretical insights into the physics of complex systems, remaining an intriguing challenge for future research.

\change{
\acknowledgments

We thank Sergey N. Dorogovtsev for stimulating discussions.
G.T. was supported by a Leverhulme Trust No. RPG-2023-187.
}

\appendix

\section{Cluster size distribution of some graph models close to criticality}
\label{AppendixA}
It is interesting to show the correctness of the scaling ansatz Eq.~\eqref{eq:scaling} in cases
for which the form of $n_s(t)$ can be exactly derived.

\subsection{$1$-d chain}
For the $1d$ chain $n_s(\phi)=\phi^2(1-\phi)$ \cite{christensen2005complexity}.
In this case $\phi_c=1$, so that $\phi=1+t$, hence 
\begin{align}
\nonumber
n_s(t)&=(-t)^2(1+t)^s
=(-t)^2\exp(s\ln(1+t))\\
&= s^{-\tau}f(s^{\sigma}t),
\end{align}
where $s_{\xi}=-1/\ln(1+t) \simeq -t^{-1} $ and $\tau=2$, $\sigma=1$.
Note that in this case $f(x) = x^2 e^{-x}$, hence $f(x)\sim x^2$ for $|x| \ll 1$.

\subsection{Erd\H{o}s-R\'enyi graphs}
Unfortunately, there are no other finite dimensional systems
for which an exact expression for $n_s(t)$ is available.
However, in Ref.~\cite{newman2007component}, Newman brilliantly derived
a general formula for the cluster size distribution in random graphs,
where he showed that
\begin{equation}
 \pi_s(\phi)=sn_s(\phi)=\frac{\phi^{s-1}\langle k \rangle}{(s-1)!}\left[\frac{d^{s-2}}{dz^{s-2}}\left[g_1(z)\right]^s \right]_{z=1-\phi}.
 \label{eq:newman_cluster_distribution}
\end{equation}
For ER graphs with $p_k=e^{-c}c^k/k!$ this implies
\begin{equation}
 \pi_s(\phi)=\frac{e^{-cs\phi}(cs\phi)^{s-1}}{s!},
\end{equation}
from which it follows, using Stirling's formula $s!~\simeq s^s e^{-s}\sqrt{2\pi s}$ for $s \gg 1$, working close to $\phi_c=1/c$, hence $c\phi=ct-1$,
\begin{align}
\nonumber
  n_s(t)
 \nonumber
 &\simeq\frac{s^{-5/2}}{\sqrt{2 \pi}} \exp(-s[ct-\ln(1+ct)])\\
 &\simeq\frac{s^{-5/2}}{\sqrt{2 \pi}} \exp(-(cs^{1/2}t)^2/2])=s^{-\tau}f(c s^{\sigma}t),
\end{align}
which is exactly in the form of Eq.~\eqref{eq:scaling}, with $f(x)$ Gaussian, and $\tau=5/2$, $\sigma=1/2$.

\subsection{Random regular graphs}
For random regular networks with degree $c>2$\footnote{The requirement
  $c>2$ is needed to avoid the case in which the network is originally
  made up entirely of loops. In this case, Newman's results, which are
  exact only for tree-like networks, fail. However, for any $\phi<1$ almost all loops break up into trees, and
  Eq.~\eqref{eq:newman_cluster_distribution} can be used, even though it is not
  well defined in the limit $\phi \to 1$.} ($c$-RRN), the generating functions are
$g_0(z)=z^{c}$ and $g_1(z)=z^{c-1}$, hence from
Eq.~\eqref{eq:newman_cluster_distribution}
\begin{align*}
\pi_s(\phi)
&=\phi^{s-1}\langle k \rangle \frac{\Gamma[s(c-1)+1]}{(s-1)!\Gamma[s(c-2)+3]}(1-\phi)^{s(c-2)+2}.
\end{align*}
Using $\Gamma(x+3)\simeq x^2\Gamma(x+1)$, and again Stirling's formula for $s\gg1$, we find, after some tedious but straightforward computations,
\begin{widetext}
\begin{align*}
 \frac{\Gamma[s(c-1)+1]}{(s-1)!\Gamma[s(c-2)+3]}&=s\frac{\Gamma[s(c-1)+1]}{s!\Gamma[s(c-2)+3]} \simeq s[s(c-2)]^{-2} \frac{[s(c-1)]^{s(c-1)}e^{-s(c-1)}\sqrt{2\pi s(c-1)}}{s^se^{-s}\sqrt{2\pi s}[s(c-2)]^{s(c-2)}e^{-s(c-2)}\sqrt{2\pi s(c-2)}}\\
 &=s^{-3/2}\sqrt{\frac{c-1}{2\pi(c-2)}}(c-2)^{-2}\exp\left[s(c-1)\ln(c-1)-s(c-2)\ln(c-2)\right]
\end{align*}
\end{widetext}
from which it follows that
\begin{align*}
 n_s(\phi)
 &\simeq \langle k \rangle \frac{s^{-5/2}}{\phi}\sqrt{\frac{c-1}{2\pi(c-2)}}\left(\frac{1-\phi}{c-2}\right)^{2}\exp\left[-s/s_{\xi}\right],
\end{align*}
where
\[s_{\xi}^{-1}=-(c-2)\ln\left[\frac{(c-1)(1-\phi)}{(c-2)}\right]-\ln[\phi(c-1)]. \]
Since $\phi_c=1/(c-1)$, we can write $c-2=(1-\phi_c)/\phi_c$, $(c-1)/(c-2)=1/(1-\phi_c)$, and $(c-1)(1-\phi)/(c-2)=1-t/(1-\phi_c)$. Therefore, expanding $\ln(1+x)$ for small $x$ and keeping the lowest orders in $t$
\begin{align*}
s_{\xi}^{-1}
&\simeq\left(\frac{1}{2\phi_c(1-\phi_c)}+\frac{1}{2\phi_c^2}\right)t^2=\frac{1}{2\phi^2_c(1-\phi_c)}t^2,
\end{align*}
from which $\sigma = 1/2$.
For $t \ll 1$, we then have
\begin{equation}
  n_s(\phi) \simeq \sqrt{\frac{(1+\phi_c)^2}{2\pi(1-\phi_c)}}s^{-5/2}e^{-s/s_{\xi}},
\end{equation}
Hence the scaling ansatz Eq.~\eqref{eq:scaling} holds, with a Gaussian scaling function, $\tau=5/2$ and $\sigma=1/2$.

\subsubsection{Explicit solution for $c=3$}
For $c=3$ it is possible to solve explicitly for $H_0(z)$, since the generating functions $G_0(z)$ and $G_1(z)$ are simple monomials of third and second order, respectively. Indeed, the equation for $H_1$,
\[H_1=z(\phi H_1 +1 -\phi)^2,\]
can be explicitly solved
\begin{equation*}
 H_1(z)=\frac{1-2z\phi(1-\phi)-\sqrt{1-4z\phi(1-\phi)}}{2z\phi^2},
\end{equation*}
which substituted in the equation
\[H_0(z)=z\left(\phi H_1(z)+1-\phi\right)^3,\]
gives
\begin{equation*}
 H_0(z)=\frac{\left[1-\sqrt{1-4z\phi(1-\phi)}\right]^3}{8z^2\phi^3}.
\end{equation*}
The convergence radius of $H_0(z)$ is $z^*=1/(4\phi(1-\phi))$. Note
that its singular behavior close to $z^*$ is determined by the
singular behavior of $H_1(z)$. The singularity is always outsite the
unitary circle apart from the critical point
$\phi=\phi_c=1/2$. Expanding $H_0(z)$ close to $z^*$, and $z^*$ for
$\phi$ close to $\phi_c$, yields $\tau=5/2$, $\sigma=1/2$.

\section{Mean-field theory for percolation and upper critical dimension}
\label{appendix:UCD}

A standard procedure in statistical physics to study the critical behavior in finite-dimensional systems is to find an equivalent, coarse-grained, description of the system of interest in terms of a field theory, i.e. a partition function
\begin{equation}
Z = \int \mathcal{D} \varphi e^{- \mathcal{S}[\varphi]}
\label{eq:partition_function}
\end{equation}
where $\mathcal{S}[\varphi]$ is
\begin{equation}
 \mathcal{S}[\varphi] = \int d^D x \left[\frac{1}{2}\varphi \Laplace \varphi +  V(\varphi)\right],
\end{equation}
$\Laplace$ is the Laplacian operator, and $V(\varphi)$ is in general a function of the ordering field
$\varphi$, whose average defines the order parameter. The potential $V$ can be determined on
the basis of symmetry considerations, and it strongly depends on the
values of the control parameters. If one finds a proper functional form for $\mathcal{S}[\varphi]$, then
the study of the critical behavior of the effective field-theory
described by Eq.~\eqref{eq:partition_function} is equivalent to the
study of the critical behavior of the finite-dimensional system of interest. For standard percolation, the effective potential contains a cubic term,~\cite{aharony1980universal,amit1976renormalization,cardy1996scaling}
\begin{equation}
 V(\varphi) = \frac{1}{2}c_2 \varphi^2 +\frac{1}{3!}c_3 \varphi^3
 \label{eq:potential_homogeneous}
\end{equation}
where $c_2 \sim \phi-\phi_c$ and $c_3$ is constant. Since in general it is not possible to solve
the integral in Eq.~\eqref{eq:partition_function}, many approximations, perturbative and non-perturbative methods have been developed in the last decades. The starting point remains, however, the so-called Landau
approximation. In this zero-th order approximation, one simply neglects
spatial fluctuations. This can be physically interpreted as a coarse-graining procedure performed over length scales larger than the correlation length $\xi$. Writing $\varphi(x)=\varphi_0 + \delta \varphi(x)$, in practice one solves the integral in Eq.~\eqref{eq:partition_function} ignoring the
contributions in $\delta \varphi(x)$. The spatial integral over $d^D x$ gives a volume
factor, hence one can compute $Z$ using the
saddle-point method, which is to find the minimum of $V(\varphi)$, $\varphi_0$ such that
\begin{equation}
 V'(\varphi_0)=0.
\end{equation}
Within this approximation, we can study the critical behavior and
compute the critical exponents. However, we must always question
whether or not the approximation is justified. We can find the answer in the Ginzburg
criterion \cite{goldenfeld2018lectures}\change{\footnote{\change{See, for instance, Ref.~\cite{bradde2010critical} for a generalization of the Ginzburg criterion to Ising model in spatial complex networks.}}}, which tells us that for a system whose Landau theory is described by a set of critical exponents $\beta, \gamma, \nu,\dots$,
the validity of the description provided by the Landau approximation is conditioned on
\begin{equation}
|t|^{-\gamma-2\beta+D\nu} \ll 1.
\end{equation}
Given $D$, this criterion provides an argument to understand whether or
not we are ``close but not too close'' to $t=0$ in order to see the
mean-field Landau exponents rather than the $D$-dependent ones. On the
other hand, this criterion tells us what is the minimum dimension $D$
of the space above which we are safe in describing our system using
the Landau approximation. This dimension is called upper critical
dimension and is given by
\begin{equation}
  D_{\text{UC}}=\frac{\gamma+2\beta}{\nu}.
  \label{Ginz}
\end{equation}
In particular, the Landau theory for standard percolation using
Eq.~\eqref{eq:potential_homogeneous} gives $\beta=1$, $\gamma=1$,
$\nu=1/2$, hence we recover the well-known result
$D_{\text{UC}}=6$ \cite{amit1976renormalization}. Since Landau theory correctly describes the critical behavior for $D \geq D_{\text{UC}}$, we can conclude that the mean-field Landau theory for
percolation, from Eq.~\eqref{eq:potential_homogeneous}, is equivalent to percolation on (infinite-dimensional) homogeneous networks. Hence to
introduce the idea of space in the infinite-dimensional world of networks we can use standard finite-dimensional results replacing $D$ with $D_{\text{UC}}$.

Homogeneous networks are effectively infinite-dimensional systems and
mean-field theory describes well the critical properties of percolation
on top of them. However, the effective field theory associated with standard percolation fails to describe the critical behavior of systems with strong heterogeneity in the connectivities and a different theory
is necessary~\cite{goltsev2003critical}.
In other words, while it is true that percolation on homogeneous networks is equivalent to the mean-field regime of percolation on finite-dimensional systems with $D>6$, percolation on heterogeneous networks obeys the mean-field theory
of a different system, in which a singular $\gamma_d$-dependent term appears
in the effective Hamiltonian~\cite{goltsev2003critical}.
In particular, it has been shown~\cite{goltsev2003critical} that the correct mean-field theory associated to percolation on heterogeneous networks with power-law degree distribution $p_k \sim k^{-\gamma_d}$ for large $k$ is given by
\begin{equation}
 V(\varphi) = \frac{1}{2}c_2 \varphi^2 +\frac{1}{3!}c_3 \varphi^3 + c_{\gamma_d} \varphi^{\gamma_d-1}.
 \label{eq:potential_heterogeneous}
\end{equation}
We can interpret Eq.~\eqref{eq:potential_heterogeneous} as the effective
potential  of a finite $D$-dimensional system with
heterogeneous connectivity of each site.
Therefore we can insert the critical exponents of the Landau theory associated to
Eq.~\eqref{eq:potential_heterogeneous}, i.e., the
exponents computed in this paper for heterogeneous networks, see Table~\ref{tab:table1}, into the
Ginzburg criterion, ~\eqref{Ginz}, to determine the upper critical
dimension $D_{\text{UC}}$. Using the values of $\beta$, $\gamma$ and $\nu$ in Table~\ref{tab:table1}, we find

\begin{equation}
 D_{\text{UC}}
 =\begin{cases}
  6,&\gamma_d \geq 4,\\
  \frac{2(\gamma_d-1)}{(\gamma_d-3)}, &3 < \gamma_d<4.
 \end{cases}
 \label{eq:UCD}
\end{equation}
\medskip

These values are in agreement with previous results in the
literature~\cite{Cohen2004} obtained via different arguments.
For $2<\gamma_d< 3$ Eq.~\eqref{eq:UCD} would predict $D_{\text{UC}}<0$.
This reflects the fact that in no finite-dimensional space,
of any arbitrary large dimension $D$, is the
critical behavior described by the Landau theory of
Eq.~\eqref{eq:partition_function} and
Eq.~\eqref{eq:potential_heterogeneous} equivalent to percolation on
heterogeneous networks.

\section{Calculations of the critical exponents for random graphs}
\label{AppendixCalculations}
In this section we develop in detail the computation of all
critical exponents for random graphs with homogeneous or power-law
degree distributions, following the general recipe described in Sec.\ref{sec:recipe}.

For homogeneous degree distributions, such as Erd\H{o}s-R\'enyi (ER) or
random regular networks,
$g_0(z)$ and $g_1(z)$ are analytic functions, hence they don't
have any singular part. In this case $b=\langle k^2 \rangle/\langle k
\rangle -1$ is the branching factor and $d=\langle k(k-1)(k-2) \rangle
/ \langle k \rangle$. The case of heterogeneous networks is provided by random
graphs with a power-law degree distribution $p_k \sim k^{-\gamma_d}$,
asymptotically for large $k$. In this case it is always possible to write\footnote{There are logarithmic corrections for
  integer $\gamma_d$.}
\begin{align}
 g_0(1-x) &\simeq 1-\langle k\rangle x +\frac{1}{2}\langle k \rangle b x^2 + C(\gamma_d-1)x^{\gamma_d-1},\\
 g_1(1-x) &\simeq 1- bx +\frac{1}{2}d x^2 + C(\gamma_d-2)x^{\gamma_d-2},
 \label{GFs}
\end{align}
where $b$, $d$ and $C(\cdot)$ depend on the value of $\gamma_d$. 
For the explicit values of these non-universal coefficients for graphs with
$p(k) = (\gamma_d-1) \km^{\gamma_d-1}k^{-\gamma_d}$, within the continuous degree approximation, see Appendix G in~\cite{cirigliano2023extended}.

\subsection{The exponents $\beta$ and $\delta$}
\subsubsection{Homogeneous degree distributions}
From Eqs.~\eqref{eq:psi_recursive} and~\eqref{g1} it follows, using $1-b\phi_c=0$, that
\begin{equation}
 -bt m +\frac{1}{2} d \phi^2 m^2 \simeq h,
 \label{eq:Psi_homogeneous}
\end{equation}
which can be solved for $m$. In particular, at criticality the linear
term vanishes and $m \sim h^{1/2}$, hence $\delta=2$ and
$E=\phi_c \langle k \rangle [2/(\phi_c^2 d)]^{1/2}$.
At $h=0$ instead, $m=0$ for $t\leq 0$ and
\begin{equation}
 m \simeq \left(\frac{2}{d\phi_c^3} \right)t,
 \label{eq:m_homogeneous}
\end{equation}
for $t>0$, hence $1-\Psi(t,0)= 2 \langle k \rangle t/(d \phi_c^2)$, i.e.,  $\beta=1$
and $B= 2 \langle k \rangle/(d \phi_c^2)$.

\subsubsection{PL degree distributions}
\label{PLbetadelta}
For $\gamma_d>4$, $d=\langle k(k-1)(k-2) \rangle$ is finite and the
singular part in $g_1(1-x)$ does not modify
Eq.~\eqref{eq:Psi_homogeneous}, which is still valid. Hence
$\beta=1$, $\delta=2$. In this case, the degree heterogeneity is not
strong enough to produce a scenario different from the homogeneous case.

For $3<\gamma_d<4$ instead, the singularity $x^{\gamma_d-2}$ in $g_1(1-x)$ (Eq.~\eqref{GFs}) is
dominant with respect to the term $x^2$ for small $x$.
Eq.~\eqref{eq:psi_recursive} becomes, using again $1-b\phi_c=0$,
\begin{equation}
 C(\gamma_d-2)\phi^{\gamma_d-2}m^{\gamma_d-2}-bt m = h,
\end{equation}
from which it follows that $m \sim h^{1/(\gamma_d-2)}$ at criticality, hence
$\delta=\gamma_d-2$, while $E=\langle k \rangle/[\phi_cC(\gamma_d-2)]^{1/(\gamma_d-2)}$.
For $h=0$ instead, $m=0$ for $t\leq 0$, while
\begin{equation}
 m \simeq \left[\frac{1}{C(\gamma_d-2)\phi_c^{\gamma_d-1}}\right]^{1/(\gamma_d-3)}t^{1/(\gamma_d-3)},
\label{m34}
\end{equation}
for $t>0$, hence $\beta=1/(\gamma_d-3)$ and
$B=\langle k \rangle [C(\gamma_d-2)\phi_c^{2}]^{1/(3-\gamma_d)}$.

The case $2<\gamma_d<3$ requires more care.
From Eq.~\eqref{GFs}, the equation for $m$ is, ($t=\phi$ since $\phi_c=0$),
\begin{equation}
 m(1+|b|t)+C(\gamma_d-2)t^{\gamma_d-2}m^{\gamma_d-2}=h,
 \label{eq:m_critical_heterogeneous}
\end{equation}
where now $b$ is a negative constant
and it is no longer the branching factor of the network (which is infinite).
Setting $t=0$, we get
$m=h$, and from Eq.~\eqref{eq:Psi_general} $\delta=1$.
Note that in this case
$\Psi(0,h)=e^{-h}\simeq 1 - h$ is analytical and no information
about the coefficients of its power series for large $s$ can be extracted
from its behavior around the origin.
Hence formally we have $\delta=1$, $E=1$, but these values are not associated to a singular behaviour, and it is no longer true that $\delta=1/(\tau-2)$,
as it will be shown below when $\tau$ is evaluated.

For $h=0$ instead
\begin{equation}
 m \simeq \left[-C(\gamma_d-2) \right]^{\frac{1}{(3-\gamma_d)}} t^{(\gamma_d-2)/(3-\gamma_d)},
\label{m23}
\end{equation}
from which, using Eq.~\eqref{eq:Psi_general}
with $\phi=t$, we get
$\beta = 1+(\gamma_d-2)/(3-\gamma_d)=1/(3-\gamma_d)$ and
$B=\langle k \rangle [-C(\gamma_d-2)]^{1/(3-\gamma_d)}$.
For $2<\gamma_d<3$ this exponent
describes the critical properties of the percolation strength for bond
percolation: $\beta_{bond}=\beta$.
As explained in the main text, for site percolation
the presence of an additional multiplicative factor $\phi$ in the equation for
$P^{\infty}$, Eq.~\eqref{eq:order_parameter}, implies
instead~\cite{radicchi2015breaking}
$\beta_{\text{site}}=1+\beta_{\text{bond}}=1+1/(3-\gamma_d)=(4-\gamma_d)/(3-\gamma_d)$.

\subsection{The exponent $\alpha$}

\subsubsection{Homogeneous degree distributions}

From Eq.~\eqref{eq:N_t_random_graphs}, expanding $\Psi(\phi,0)$ for small
$t$, using $m=m(t,0)$ as in Eq.~\eqref{eq:m_homogeneous} and $b\phi-1=bt$ we get
\begin{equation}
  F(t)\simeq 1 -\frac{\langle k \rangle \phi_c}{2}-\frac{\langle k \rangle }{2}t
  +\frac{\langle k \rangle}{2}tm^2-\frac{1}{3!}\langle k \rangle d \phi_c^3 m^3.
\label{Nt}
\end{equation}

Eq.~\eqref{Nt} shows that $F(t)$ approaches a constant
value linearly in $t$ as $t \to 0$. However, the exponent $\alpha$
accounts for the behavior of the singular part of $F(t)$,
which is given by the terms involving $m$.
For homogeneous distributions, since $m \sim t$ for $t>0$
[see Eq.~\eqref{eq:m_homogeneous}] we get
$F(t)\sim t^{3}$, hence $\alpha=-1$, in agreement with the
scaling relations.
In this case, the exponent governing the behavior of the singular part
is, by chance, an integer.
As shown below, this is not the case for PL degree distributions.

Note also that, for any $t<0$, $m$ is identically zero.
This means that there is no singular contribution in $F(t)$
for $t<0$. Hence the exponent $\alpha$ and the associated amplitude
$A_-$ are not defined in the nonpercolating phase. For this reason, we did not consider in our analysis two additional amplitude ratios containing $A_-$ defined in \cite{aharony1980universal}.

\subsubsection{PL degree distributions}

From Eq.~\eqref{eq:N_t_random_graphs}, expanding $\Psi(\phi,0)$ for small
$t$ and using $m=m(t,0)$ we find,
neglecting the non-singular term $1-\phi \langle k \rangle /2$,
\begin{equation}
  F(t)\simeq \frac{\phi \langle k \rangle}{2}
  (b\phi-1)m^2-\frac{\langle k \rangle d}{3!} \phi^3
m^3+C(\gamma_d-1)\phi^{\gamma_d-1}m^{\gamma_d-1}.
\end{equation}

For $\gamma_d > 4$, the singular term proportional to $m^{\gamma_d-1}$
can be neglected. Since $m \sim t$ and $b\phi-1=bt$,
as for homogeneous degree distributions we get $\alpha=-1$.

For $3<\gamma_d < 4$, the only difference is that
$m \sim t^{1/(\gamma_d-3)}$ [see Eq.~\eqref{m34}],
hence at leading singular order
\begin{equation}
 F(t) \sim t^{(\gamma_d-1)/(\gamma_d-3)},
\end{equation}
from which $\alpha=(\gamma_d-5)/(\gamma_d-3)$.

For $2<\gamma_d<3$, $\phi=t$, and
$m \sim t^{(\gamma_d-2)/(3-\gamma_d)}$ [see Eq.~\eqref{m23}].
Now the leading singular term is the one proportional to $t^{\gamma_d-1}m^{\gamma_d-1}$
(since $\gamma_d-1<2$), so that
\begin{equation}
F(t) \sim t^{(\gamma_d-1)/(3-\gamma_d)},
\end{equation}
from which
$\alpha=(7-3\gamma_d)/(3-\gamma_d)$.

Again the exponent $\alpha$ and the associated amplitude $A_-$ are not defined
in the nonpercolating phase.

\subsection{The exponent $\gamma$}

\subsubsection{Homogeneous degree distributions}

For $t<0$, at $h=0$ we have $m=0$.
Hence from Eq.~\eqref{sminus1}
\[\langle s \rangle-1 \simeq \frac{\phi_c \langle k \rangle}{-bt} = \phi_c^2 \langle k \rangle |t|^{-1}, \]
giving $\gamma = 1$ and $C_{-}=\phi_c^2 \langle k \rangle$.
As expected, $C_{-}$ is a nonuniversal constant, since it depends on the
details of the model. For $t>0$ instead, $m$ is given by
Eq.~\eqref{eq:m_homogeneous}, and
\[\langle s \rangle-1 \simeq \frac{\phi_c \langle k \rangle }{-t b +2bt } = \phi_c^2 \langle k \rangle  t^{-1}, \]
from which again $\gamma=1$ and the (universal) amplitude ratio is $C_{+}/C_{-}=1$.

\subsubsection{PL degree distributions}

For $\gamma_d>4$, the results obtained for homogeneous distributions
hold.

For $3<\gamma_d<4$, still $m=0$ for $t<0$, from which we get again the same behavior approaching the threshold from below.
For $t>0$ instead, inserting Eq.~\eqref{m34} into Eq.~\eqref{sminus1} we have at leading order
\begin{align*}
  \langle s \rangle -1&
\simeq \frac{\phi_c \langle k \rangle}{1-\phi\big[b-(\gamma_d-2)C(\gamma_d-2)\phi_c^{\gamma_d-3}m^{\gamma_d-3}\big]}\\&=\frac{\phi_c \langle k \rangle}{-bt+(\gamma_d-2)bt}=\frac{\phi_c^2 \langle k \rangle}{(\gamma_d-3)}t^{-1}.
\end{align*}
In this case the exponent is $\gamma=1$ and
the critical amplitude ratio is $C_{+}/C_{-}=1/(\gamma_d-3)$.

For $2<\gamma_d<3$, inserting Eq.~\eqref{m23} into Eq.~\eqref{sminus1} at leading order we get
\begin{align*}
\langle s \rangle -1 &\simeq \frac{t\langle k \rangle}{1-t\big[-|b|-(\gamma_d-2)C(\gamma_d-2)t^{\gamma_d-3}m^{\gamma_d-3} \big]}\\
&\simeq \frac{\langle k \rangle}{3-\gamma_d} t.
\end{align*}
This implies $\gamma=-1$ and $C_+=\langle k \rangle/(3-\gamma_d)$.

\subsection{The exponent $\nu$}

\subsubsection{Homogeneous degree distributions}

Close to criticality, taking $\phi<\phi_c$, since $u=1$ and $g_1'(1)=b$, from
Eq.~\eqref{xi2} it follows $\xi \simeq (1-\phi b)^{-1/2}=b^{-1/2} (-t)^{-1/2}$.
If $t>0$, we have $u=1-m$ and $g_1'(1-\phi m)$ can be expanded using Eq.~\eqref{eq:m_homogeneous}
obtaining $\xi \simeq (\phi b-1)^{-1/2}=b^{-1/2} t^{-1/2}$.
In both cases the prefactor is the same so that
the amplitude ratio $\Xi_+/\Xi_-=1$.

\subsubsection{PL degree distributions}

For any $\gamma_d>3$, since $b$ is finite an analogous argument also applies
yielding $\nu=1/2$ and $\Xi_+/\Xi_-=1$.
When $2<\gamma_d<3$, instead, since $\phi=t$, the denominator in Eq.~\eqref{xi2}
goes to a constant, while the numerator is of order $t$.
As a consequence, $\xi \sim t^{1/2}$ and hence $\nu=-1/2$.

\subsection{The amplitude ratio $R_\chi$}

\subsubsection{Homogeneous degree distributions}
Inserting in the definition
\begin{equation}
R_\chi = C_+ E^{-\delta} B^{\delta-1}
\label{eq:R_chi_def}
\end{equation}
the value $\delta=2$ and the expressions for $C_+$, $E$ and $B$ derived above
for homogeneous distributions yields
\begin{equation}
R_\chi=1,
\end{equation}
the same universal value obtained on lattices above the upper critical
dimension~\cite{aharony1980universal}.

\subsubsection{PL degree distributions}
\label{RchiPL}
For $\gamma_d>4$ the amplitudes are the same of the homogeneous case
so that $R_\chi=1$.  For $3<\gamma_d<4$ instead, $\delta=\gamma_d-2$
and the expressions of the amplitudes $C_+$, $B$ and $E$ are different
(see Sec.~\ref{PLbetadelta}).  Inserting them into the definition
yields $R_\chi=1/(\gamma_d-3)$, which has the same level of
universality of the exponents, i.e., it depends only on $\gamma_d$ but
not on details of the degree distribution.  For $2<\gamma_d<3$, since
$\delta=1$ and $E=1$ we find $R_\chi=C_+=\langle k \rangle/(3-\gamma_d)$.
Note that this value is not universal (it
depends on $\av{k}$), as it should be. However, we already pointed out
that the point $t=0$ for $2<\gamma_d<3$ is not a true critical point,
and the exponent $\delta=1$, together with the amplitude $E=1$, coming
from the analytic part of $\Psi$ are not associated to a critical
behavior. However, a nonrigorous argument can be developed to restore
criticality and define a universal $R_{\chi}$ even in this case. The
argument is as follows. The singular part of $\Psi(t,h)$ can be
written, from Eq.~\eqref{eq:Omega_scaling_form}, as
$\{ \Psi(t,h) \}_{\text{sing}}=|t|^{\frac{\tau-2}{\sigma}} {\mathcal F}_\pm'(x)$,
where ${\mathcal F}_\pm'$ is a scaling function of the variable
$x=h/|t|^{1/\sigma}$. The critical behavior -- the exponents $\beta$,
$\delta$ and the associated critical amplitudes -- can be recovered
from the cases $x \ll 1$, i.e. $h \to 0$ and $t$ small, and $x \gg 1$,
i.e. $t \to 0$ and $h$ small. For $\gamma_d>3$ as well as for
homogeneous degree distributions, ${\mathcal F}_\pm'(x)\sim 1$ for $x \ll 1$ and
${\mathcal F}_\pm'(x)\sim x^{\tau-2}$ for $x\gg 1$. However, for $2<\gamma_d<3$, as no
singular part of $\Psi$ remains when $t \to 0$, the scaling behavior
of ${\mathcal F}_\pm'(x)$ for $x \gg 1$ must be different, and must correspond to $\{\Psi(t\to 0,h)\}_{\text{sing}} \sim 0$.
Working instead at $x=\mathcal{O}(1)$,
i.e., small but finite $h$ and $t$, from
Eq.~\eqref{eq:m_critical_heterogeneous} it follows, using the value of
$\sigma$ in Table~\ref{tab:table1}, at lowest order $m\simeq
[-C(\gamma_d-2)]^{1/(3-\gamma_d)}x^{-1}h$. Plugged into
Eq.~\eqref{eq:Psi_general}, with $\phi=t=x^{-\sigma}h^{\sigma}$, we
finally get
\[\Psi(\phi,h) \simeq 1 - \av{k}[-C(\gamma_d-2)]^{\frac{1}{(3-\gamma_d)}}x^{\frac{1}{2-\gamma_d}}h^{1/(\gamma_d-2)}, \]
from which $\delta'=1/(\gamma_d-2)$ and $E'=\av{k}[-C(\gamma_d-2)]^{\frac{1}{(3-\gamma_d)}}x^{\frac{1}{2-\gamma_d}}$. Using these values, from Eq.~\eqref{eq:R_chi_def}, we get
\[R_{\chi}=\frac{1}{3-\gamma_d}[-C(\gamma_d-2)]^{\frac{1}{\gamma_d-3}}x. \]
Fixing $x=[-C(\gamma_d-2)]^{\frac{1}{3-\gamma_d}}$ we then find the universal value $R_{\chi}=1/(3-\gamma_d)$. Thus for $2<\gamma_d<3$ we can formally study the singular behavior of $\Psi$ in the limit $t \to 0$, $h \to 0$ but keeping $x$ fixed as above. Note that $\delta'$ is also in agreement with the scaling relation $\tau=2+1/\delta '$.
The physical motivation for the choice of $x$ remains to be understood.

\subsection{The exponents $\sigma$ and $\tau$}

\subsubsection{Homogeneous degree distributions}
Using Eq.~\eqref{eq:psi_homogeneous} and~\eqref{eq:dpsi_homogeneous}
in Appendix~\ref{appendix:psi_asymptotic}, the solution of the characteristic
equation is $\eta=t/(d \phi_c^3)$.
Hence, since $u^*=1-\eta$
\begin{equation}
\rho = \psi(u^*) \simeq 1 + \frac{1}{d \phi_c^4}t^2,
\label{eq:rho_homogeneous}
\end{equation}
from which, using
the definition of $s_\xi$ in Eq.~\eqref{eq:definition_s_xi}, $\sigma=1/2$.
For $t>0$, $\rho$ is far away from the unitary disk: the presence of the GC strongly
hinders the formation of other large clusters, introducing an exponential
cutoff~\cite{newman2007component}.
At criticality $\phi=\phi_c$,
$t=0$, the singularity $\rho$ touches the unitary disk
$\rho=1$, and $s_\xi$ diverges.
For $t<0$, in the phase without a GC, the singularity $\rho$ is again pushed
away from the unitary circle since, also in this case,
only finite clusters can form.

In order to derive $\tau$ we must invert Eq.~\eqref{inversion}.
From the expansion of $\psi(u)$ around $u^*$, setting $\epsilon=u^*-u$,
see Eq.~\eqref{eq:taylor_psi}, we obtain
\begin{align*}
\rho-z &=\psi(u^*)-\psi(u)\simeq\frac{1}{2}|\psi''(u^*)|\epsilon^2,
\end{align*}
which can be inverted\footnote{The inversion produces an
  ambiguity in the determination of the sign. The sign
  is taken
  in order to have an increasing $H_1(z)$ for $z \to \rho^-$.}
to give $\epsilon \sim (\rho-z)^{1/2}$, yielding
\begin{equation}
 H_1(z) \simeq u^* - A_1(\rho-z)^{1/2},
\end{equation}
with $A_1$ a positive constant. This result implies
\begin{equation}
  \pi_s(\phi) \simeq q_0 s^{-3/2}e^{-s/s_{\xi}}=q_0 s^{-3/2}f(q_1 s^{1/2} t),
  \label{pi_homo}
\end{equation}
from which we find that $\tau=5/2$ and the universal scaling
function is Gaussian $f(x)=e^{-x^2}$.  As noted
in~\cite{kryven2017general}, this behavior is a consequence of the
central limit theorem.
Note that Eq.~\eqref{pi_homo} holds for any $|t| \ll 1$ (see Fig.~\ref{Figsummary}(a)).

\subsubsection{PL degree distributions}

For power-law degree-distributed networks it is necessary to analyze
separately what happens above and below the threshold and consider
various ranges of $\gamma_d$ values.

\paragraph{$\gamma_d>4$}

\subparagraph{$t>0$:}
For positive $t \ll 1$, the expansions of $\psi(u^*)$ and $\psi'(u^*)$
in Eqs.~\eqref{eq:psi_heterogeneous},~\eqref{eq:dpsi_heterogeneous} are
equivalent to
Eqs.~\eqref{eq:psi_homogeneous},\eqref{eq:dpsi_homogeneous}, hence we
recover again Eq.~\eqref{eq:rho_homogeneous}, and
$\sigma=1/2$.
Furthermore, from Eq.~\eqref{eq:taylor_psi} we get for $s \gg 1$
\begin{equation}
 n_s(t) \sim s^{-5/2}e^{-s/s_\xi}.
 \label{eq:pi_s_tail_t>0_gamma_4}
\end{equation}

\subparagraph{$t=0$:}
At criticality $\psi(u)$ is singular for $u=1$ and its expansion
is now given by Eq.~\eqref{eq:asymptotic_psi_t=0}.
However, since $\epsilon^{\gamma_d-2}\ll \epsilon^2$, we get the same asymptotic
behavior as in Eq.~\eqref{eq:pi_s_tail_t>0_gamma_4}, even though
$\gamma_d$-dependent subleading corrections are present.
We can conclude that $\tau=5/2$.

\subparagraph{$t<0$:}
In this case the convergence radius of $H_1(z)$
remains $\rho=1$, since the characteristic equation $\psi'(u^*)=0$
admits a solution only for $t>0$. Hence $s_\xi=\infty$. In the expansion of $\psi(u)$ around 1 (see Eq.~\eqref{eq:asymptotic_psi_t<0})
if the condition
\begin{equation}
  -t \ll \epsilon
  \label{condition}
\end{equation}
holds, then the term proportional to $\epsilon$ can be
neglected compared to the term $\epsilon^2$. In such a case
the inversion of Eq.~\eqref{eq2:asymptotic_psi_t<0} yields
\begin{equation}
  \epsilon \simeq (1-z)^{1/2}.
  \label{eps}
\end{equation}
Inserting this
into Eq.~\eqref{condition} consistency is found as long as
$1-z \gg (-t)^2 = (-t)^{1/\sigma}$.
Hence $n_s$ decays with exponent $\tau=5/2$
for $s \ll s_0 \sim (-t)^{-1/\sigma}$.
When this condition is not fulfilled, then at leading order from
Eq.~\eqref{eq2:asymptotic_psi_t<0} $\epsilon \sim (1-z)$ which,
replaced back into Eq.~\eqref{eq2:asymptotic_psi_t<0}, gives
\begin{equation}
  1-H_1(z) = \epsilon = c_1(1-z)+\textrm{other reg. terms} + c_2(1-z)^{\gamma_d-2}.
  \label{eq3:asymptotic_t<0n}
\end{equation}
Since the first term in the r.h.s. is regular, 
this expression implies, via Eq.~\eqref{pi_tauberian}, that
\begin{equation}
n_s(t) \sim s^{-\gamma_d}.
\label{gammad}
\end{equation}

In summary, we find the critical decay $n_s \sim s^{-5/2}$
for $1\ll s \ll s_0$, with $s_0 \sim |t|^{-1/\sigma}$;
for $s \sim s_0$,
$n_s(t)$ exhibits an exponential cutoff and then, for $s \gg s_0$, a tail
$s^{-\gamma_d}$ (see Fig.~\ref{Figsummary}(b)).

Hence, even though the correlation size $s_\xi$ is formally infinite for $t<0$, there is another
characteristic size $s_0$ such that the cluster size distribution can
be written as
\begin{equation}
 n_s(t) \simeq q_0s^{-\tau}f_{-}(q_1s^{\sigma}t) + q_2 s^{-\gamma_d},
 \label{eq:scaling_t<0}
\end{equation}
with $f_{-}(x) \ll 1$ for $|x| \ll 1$, showing that the scaling ansatz
Eq.~\eqref{eq:scaling} must be modified.

Two distinct mechanisms contribute to the cluster size distribution
tail in Eq.~\eqref{eq:scaling_t<0}.
The first is the collective phenomenon of percolation, where
finite clusters merge into larger and larger clusters as $t \to 0^-$.
This mechanism is responsible for the critical properties
such as the divergence of $\langle s \rangle $. The second is the
presence for any finite $\phi>0$
of a small but finite fraction of nodes with arbitrarily high degree,
surrounded by active neighbors, as already noticed in~\cite{newman2007component}.
The contribution of the first mechanism is larger than the second,
all the more so when criticality is approached, 
because $\gamma_d>\tau$, so that only the first tail remains
as $s_0$ diverges.
Therefore, even if Eq.~\eqref{eq:scaling} is not strictly obeyed,
the correction in Eq.~\eqref{eq:scaling_t<0} can be neglected if we are
interested in critical properties.

\vspace{0.5cm}

\paragraph{$3<\gamma_d<4$}

\begin{figure}[h!]
    \includegraphics[width=0.98\columnwidth]{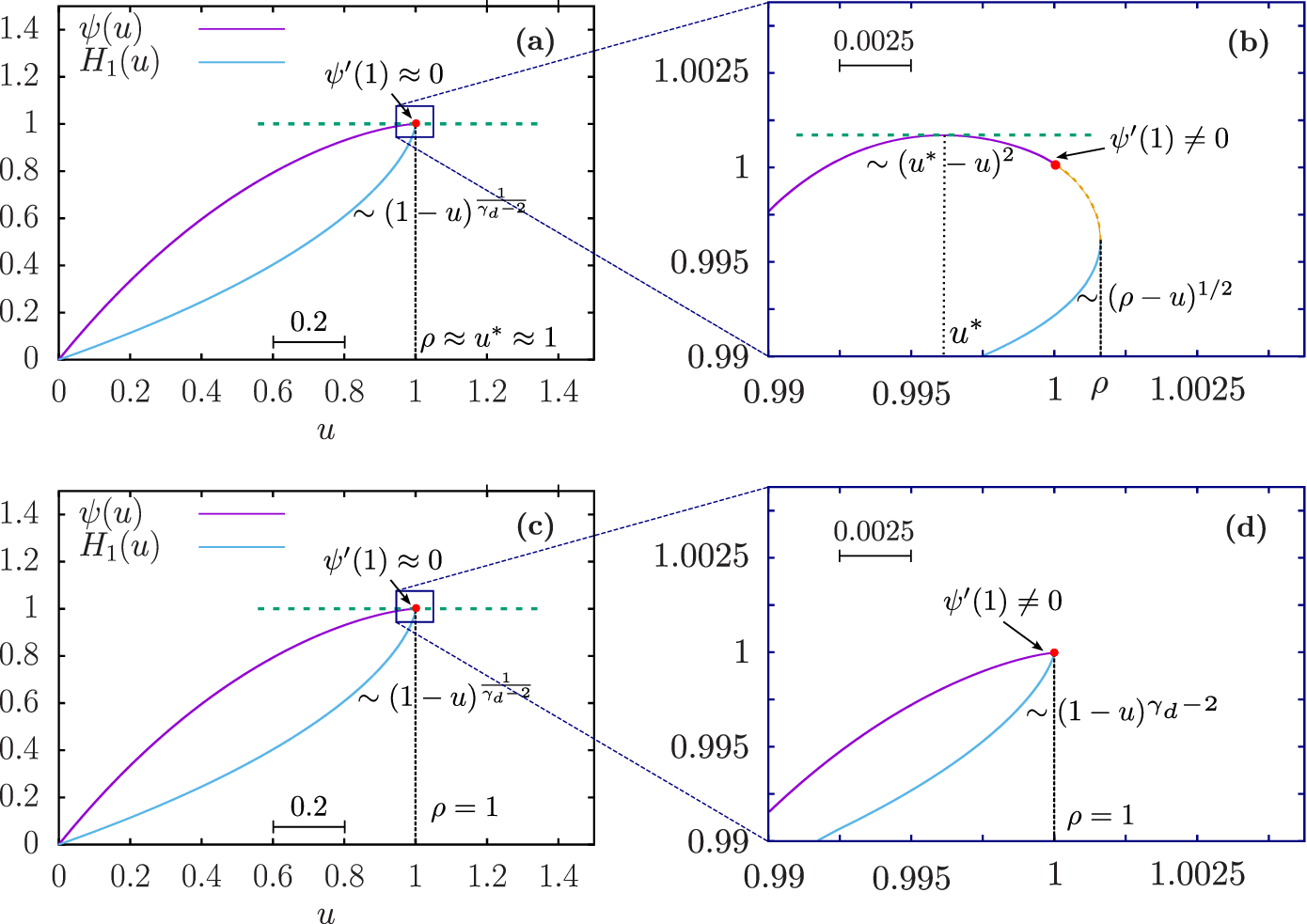}
    \caption{The origin of the crossovers in the cluster size distributions for power-law networks. In this case, $\gamma_d=3.5$ is considered. In (a), the functions $\psi(u)$ and its inverse $H_1(u)$ are plotted for $t>0$ but $t\ll 1$. In particular, on the scale in (a) it is impossible to detect the difference between $\rho$, $u^*$ and $1$, and the system seems to be at criticality $t=0$ and $\rho=1$, with a singular behaviour associated with the critical exponent $\tau$. On a closer look (b), however -- hence on a smaller scale -- we can distinguish $u^*$ from $1$ and a square-root singularity determines the behaviour close to the singularity at $\rho>1$. The asymptotic scaling of $n_s$ for large $s \gg s_*$ is then determined by the inversion in (b), but for $1 \ll s \ll s_*$ the picture in (a) is valid and determines the preasymptotic scaling with exponent $\tau$. Similarly, in (c) the situation $t<0$ but $|t|\ll 1$ is considered. At first sight, the system seems to be at criticality $t=0$ in (c), where $\psi'(1)=0$. However, on a closer look (d) we realize that $t<0$ since $\psi'(1) \neq 0$. Again, the picture in (c) determines the scaling of $n_s$ for $1\ll s \ll s_0$, while asymptotically we must look closer (d) to find the scaling $n_s \sim s^{-\gamma_d}$ for $s \gg s_0$.}
    \label{Figzoom}
\end{figure}

\subparagraph{$t>0$:}
For positive $t \ll 1$, in
Eq.~\eqref{eq:dpsi_heterogeneous}
the subleading term proportional to $\eta$ can be neglected.
Inserting the solution
of the characteristic equation into Eq.~\eqref{eq:psi_heterogeneous}
we obtain
\begin{equation}
\rho \simeq 1 +\frac{\phi_c^{-(\gamma_d-2)/(\gamma_d-3)}}{\left[(\gamma_d-2)C(\gamma_d-2)\right]^{1/(\gamma_d-3)}}t^{(\gamma_d-2)/(\gamma_d-3)},
\end{equation}
from which $\sigma=(\gamma_d-2)/(\gamma_d-3)$.
Since $t>0$, Eq.~\eqref{eq:taylor_psi} can be used again,
from which we find again the scaling in Eq.~\eqref{eq:pi_s_tail_t>0_gamma_4}.
However, now Eq.~\eqref{eq:taylor_psi} is valid conditioned on
$\epsilon \ll 1-u^*=\eta$,
because of the presence of the singularity of $G_1(u)$ in $u=1$\footnote{Note that such a singularity is present also for $\gamma_d>4$,
but in that case the leading term produced by the inversion is still given by a square root singularity, as already noted above. The singular behaviour close to and at criticality are the same and no crossover is observed.}.
If we consider the inverse function $H_1(z)$ (see Fig.~\ref{Figzoom}) the inversion works only
far from the singularity in $z=1$, which implies
$\rho-z \ll \rho-1 = \psi(u^*)-1 \ll 1$.
Hence
the asymptotic behavior obtained
from Eq.~\eqref{eq:taylor_psi} holds provided $s \gg s_*$, where
\begin{equation}
 s_*=\frac{1}{\psi(u^*)-1}.
\end{equation}
In the opposite regime $\psi(u^*)-1 \ll \rho-z \ll 1$, that is for
$1 \ll s \ll s_*$, it is not possible to distinguish $u^*$ from $1$, hence
$\rho$ from $1$, and the validity of the Taylor expansion
breaks down.
We can use instead the asymptotic expansion close to $1$
-- see Eq.~\eqref{eq:asymptotic_psi_t=0} -- rather than the Taylor expansion close
to $u^*$. In this regime, we observe the critical behavior, see the case $t=0$ below. Note that since $s_\xi = 1/\log(\rho)=1/\log(1+\psi(u^*)-1)$,
and $\log(1+x) \leq x$, we have $s_* \leq s_\xi$.
In particular, $s_* \sim s_\xi$ for $t \ll 1$.

\subparagraph{$t=0$:}
In this case Eq.~\eqref{eq:asymptotic_psi_t=0} must be used, where now
$\epsilon^2 \ll \epsilon^{\gamma_d-2}$.
Inverting Eq.~\eqref{inversion}
implies $\epsilon \sim (\rho-z)^{1/(\gamma_d-2)}$
which leads to
\begin{equation}
  n_s(0) \sim s^{-[2+1/(\gamma_d-2)]},
\end{equation}
i.e., $\tau=2+1/(\gamma_d-2)$.

There is an apparent contradiction: the asymptotic scaling with
exponent $5/2$ for any $t>0$ is different from what occurs for $t=0$,
where $\tau=2+1/(\gamma_d-2)>5/2$.  This is at odds with what happens
for homogeneous systems (and for $\gamma_d>4$) where the value of the
exponent is the same and only the cutoff scale $s_\xi$ changes as
criticality is approached.  The conundrum is solved by the presence of
the crossover scale $s_*$ (discussed above for $t>0$), which separates
a decay as $s^{-\tau}$ for $1 \ll s \ll s_*$ from the asymptotic decay
$s^{-5/2} e^{-s/s_\xi}$.  When $t \to 0$ the crossover size diverges,
$s_* \sim t^{-1/\sigma}$, while still remaining smaller than the
correlation size $s_* \leq s_{\xi}$
This does not imply a violation of the scaling ansatz.
Eq.~\eqref{eq:scaling} is still valid but, analogously to
what happens
in the case $2<\gamma_d<3$ (see below), the scaling function $f_+(x)$
has a nonstandard form: it is constant for $x \ll x^*=s_*/s_{\xi}$;
it vanishes exponentially for $x \gg 1$; it behaves as $x^{(4-\gamma_d)/[2(\gamma_d-3)]}$ in the interval between $x^*$ and 1.

\subparagraph{$t<0$:}
The phenomenology is perfectly analogous to the
case $\gamma_d>4$: $n_s(t)$ decays as $s^{-\tau}$ for $1 \ll s \ll s_0$,
where $s_0$ diverges as $|t|^{-1/\sigma}$ as criticality is
approached. For $s \gg s_0$ instead $n_s(t) \sim s^{-\gamma_d}$.  The
only difference with respect to the case $\gamma_d>4$ is that now
$\tau=2+1/(\gamma_d-2)$, because the term competing with the linear
term in Eq.~\eqref{eq:asymptotic_psi_t<0} is $\epsilon^{\gamma_d-2}$
instead of $\epsilon^2$.

\vspace{0.5cm}

\paragraph{$2<\gamma_d<3$}

\subparagraph{  }

In this range of $\gamma_d$ values, only the phase $t>0$ exists and $\phi=t$.
We can solve again the characteristic equation using the expansion
Eq.~\eqref{eq:dpsi_heterogeneous} for small $\eta$,
and from Eq.~\eqref{eq:psi_heterogeneous} obtain
\begin{equation}
 \rho =1-C(\gamma_d-2)[-(\gamma_d-2)C(\gamma_d-2)]^{1/(3-\gamma_d)}t^{(\gamma_d-2)/(3-\gamma_d)},
\end{equation}
from which $\sigma=(3-\gamma_d)/(\gamma_d-2)$.
This result implies
the existence of a diverging correlation size $s_{\xi}\sim t^{-1/\sigma}$
even if for $t=0$ no critical behavior occurs.

The asymptotic decay of $n_s(t)$ is again
determined by the singularity produced by the inversion of
$\psi(u)$ close to $u^*$
\begin{equation}
 H_1(z) \simeq u^{*}-c_1|\psi''(u^*)|^{-1/2}(\rho-z)^{1/2}.
\end{equation}
Hence $n_s$ exhibits a tail $s^{-5/2}$ with an exponential cutoff at
$s_{\xi}$.
As in the case $3<\gamma_d<4$, this
expression holds provided that $u$ is close enough to $u^*$, hence the
asymptotic behavior extracted from it holds only for
$s \gg s_* \sim t^{-1/\sigma}$.
For $1 \ll s \ll s_*$, we can use the expansion in
Eq.~\eqref{eq:asymptotic_psi_general}, in perfect analogy to the cases
of PL networks for $t<0$ considered above, which gives
\begin{equation}
H_1(z) \simeq 1 -c_2 \phi^{\gamma_d-2}(1-z)^{\gamma_d-2},
\end{equation}
corresponding to a power law scaling of $n_s$ with exponent
$\gamma_d$. Following the analogy with the case $3<\gamma_d<4$, we may
be tempted to identify $\tau=\gamma_d$: as $t \to 0$, only the decay
with exponent $\gamma_d$ survives.
However it has to be noted that there are multiplicative factors
depending on $\phi$.
While they do not play any essential role for $\gamma_d>3$, as
for $t \ll 1$ they are simply $\phi_c >0$ at leading order, the
situation now requires a more careful analysis.
From Eq.~\eqref{eq:asymptotic_psi_general},
and the expression of $\eta=1-u^*$ in terms of $t$, we see that
$|\psi''(u^*)| \sim t ^{(\gamma_d-3)/(\gamma_d-2)}=t^{-1/\sigma}$.
The concavity of
$\psi(u)$ in $u^*$ diverges as $t \to 0$.
This means that the peak around $u^*$ gets not only smaller,
but also narrower, as $t \to 0$.
This reflects the fact that $\psi(u)=u$ is analytic
for $\phi=0$ and there is no peak at all. Hence a prefactor
proportional to $t^{1/\sigma}$ multiplies the decay of $n_s$ for large $s$.
Similarly, the preasymptotic scaling for $1\ll s \ll s_*$
contains a prefactor $t^{\gamma_d-2}$.
Finally, an additional factor $t$ in $n_s(t)$ comes from the mapping between the
singular behavior of $H_1$ and the singular behavior
of $H_0$, Eq.~\eqref{H0sing}, since
$B_0=\rho G_0'(u^*) A_1 \sim t$.
Putting all these results together
\begin{align}
 n_s(t) \simeq \begin{cases}
               c_a t^{\tilde{a}}s^{-\gamma_d},& 1 \ll s \ll s_*,\\
               c_b t^{\tilde{b}}s^{-\frac{5}{2}}e^{-s/s_{\xi}},& s \gg s_*,
              \end{cases}
            \label{eq:n_s_scalefree1}
\end{align}
where $s_\xi \sim s_* \sim t^{-1/\sigma}$, $\sigma=(3-\gamma_d)/(\gamma_d-2)$,
$\tilde{a}=\gamma_d-1$, $\tilde{b}=1/(2\sigma)+1$, and $c_a,c_b$ are constants
independent of $t$.
The expression in Eq.~\eqref{eq:n_s_scalefree1} can be cast in a scaling form like Eq.~\eqref{eq:scaling}, in which the dependence on $t$ is absorbed in the scaling function. Hence
\begin{align}
 n_s(t)\simeq s^{-\tau} f_+(q_1 s^{\sigma}t),
 \end{align}
where $\tau=2+1/(\gamma_d-2)$ and
the scaling function $f_+(x)$ behaves now as
\begin{align}
      f_+(x) \simeq
    \begin{cases}
      (c_a/q_1^{\tilde a})x^{\tilde{a}},&   x \ll x_*,\\
      (c_b/q_1^{\tilde b})x^{\tilde{b}}e^{-x^{1/\sigma}}, & x \gg x_*,
    \end{cases}
    \label{eq:scaling_function_strange}
\end{align}
with $x_* = q_1 \widetilde{q}$.
Note that if $x_*>1$, only the first scaling,
followed by the exponential cutoff for $x \sim 1$, is observable.
This form of the scaling function with $f_+(x) \to 0$ as $x \to 0$
is unusual, but
note that a similar behavior of the scaling function appears also for
the linear chain (see Appendix~\ref{AppendixA}),
where $n_s$ can be put in a scaling form like Eq.~\eqref{eq:scaling}
with $\tau=2$ and $f_-(x) \sim x^2 \to 0$ for $x \to 0$.

In conclusion, the analysis developed here shows that 
for $2<\gamma_d<3$ the cluster size distribution obeys the scaling form in
Eq.~\eqref{eq:scaling} with $\sigma=(3-\gamma_d)/(\gamma_d-2)$,
$\tau=2+1/(\gamma_d-2)$ but with a scaling function $f_+(x) \sim x^{\gamma_d-1}$
for $x \to 0$.
This implies that if the decay of $n_s(t)$ with $s$ is studied at fixed $t$
(as it is usually the case) the exponent measured is not $\tau$ but $\gamma_d$ with
a prefactor decreasing as $t \to 0$.

\subsection{The exponent $\bar \nu$}
\label{appendix:bar_nu}

For power-law degree-distributed networks with $\gamma_d>3$,
$b(k_c) \simeq b_\infty - C_2 k_c^{-(\gamma_d-3)}$,
where $b_\infty=b(k_c=\infty)$ and $C_2$ is a positive constant.
Inserting this expression into Eq.~\eqref{phic0} we obtain, at lowest
order,
\begin{equation}
  \phi_c(N,k_c) \simeq b_\infty^{-1} + \frac{C_2}{b_\infty^2} k_c^{-(\gamma_d-3)} +
  b_\infty^{-1} C_1 N^{-1/3}.
\label{phic}
\end{equation}

If degrees are sampled from a distribution with a hard (structural) cutoff
$\kmax(N) \sim N^{1/\omega}$ the actual cutoff in a realization of
the network is the minimum between the structural cutoff and the natural one,
$N^{1/(\gamma_d-1)}$. This implies that in full generality
$k_c(N) \sim N^{1/{\tilde \omega}}$
with ${\tilde \omega}=\max[\omega,\gamma_d-1]$~\footnote{The
  exponent ${\tilde \omega}$ must be larger than or equal to $2$,
  otherwise the network is necessarily
  correlated~\cite{catanzaro2005generation}}.

Then, considering Eq.~\eqref{phic}, the scaling
of the effective threshold depends on whether $(\gamma_d-3)/{\tilde \omega}$
is larger or smaller than $1/3$.

For ${\tilde \omega} \le 3(\gamma_d-3)$
the second to last term in Eq.~\eqref{phic} decays
faster than the last one;
as a consequence $\bar{\nu}=3$ as in the fully homogeneous case:
the scaling of the effective threshold $\phi_c(N)$ is governed by
bona fide critical fluctuations. This occurs for $\gamma_d \ge 4$
and $\omega \le 3(\gamma_d-3)$.

\begin{figure}[h!]
    \includegraphics[width=\columnwidth]{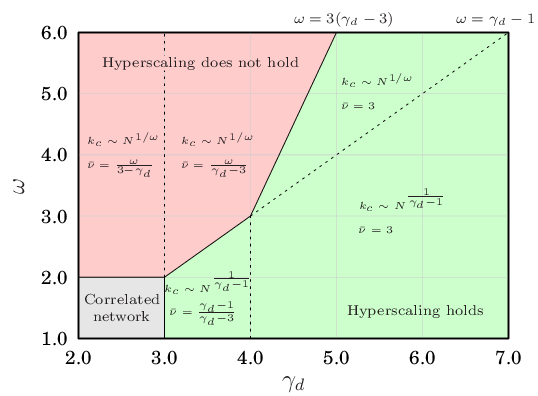}
    \caption{Finite size scaling behavior as a
      function of $\gamma_d$ and $\omega$.}
    \label{Fig_nubar}
\end{figure}

If instead ${\tilde \omega} \ge 3(\gamma_d-3)$ the role of the two contributions
is reversed: in this case, for asymptotically large $N$ the second to
last contribution in Eq.~\eqref{phic} dictates the scaling of
$\phi_c(N)$, yielding ${\bar \nu} = {\tilde \omega}/(\gamma_d-3)$.
For $3 < \gamma_d < 4$ and $\omega \le \gamma_d-1$
(so that ${\tilde \omega}=\gamma_d-1$) this implies
${\bar \nu} = (\gamma_d-1)/(\gamma_d-3)$.
Otherwise ${\tilde \omega}=\omega$ and ${\bar \nu} = \omega/(\gamma_d-3)$, see Fig.~\ref{Fig_nubar}.
The fact that $\bar \nu$ depends on the parameter $\omega$, which
governs how the thermodynamic limit is reached, immediately implies
that the hyperscaling relation~\eqref{eq:hyperscaling} is violated
as the left hand side cannot depend on $\omega$.
The violation of~\eqref{eq:hyperscaling} can be traced back to
the breakdown of Eq.~\eqref{thetahs}.
In its turn this is interpreted as follows: the slow growth of
the maximum degree $k_c$ with $N$ introduces an additional cutoff
smaller than $s_\xi$ in the cluster size distribution.

For $2 < \gamma_d <3$ the picture is completely different.
The argument based on the mapping to an appropriate field-theory
puzzlingly yields a negative upper critical dimension.
It is more useful to observe that the negative value of
the exponent $\nu=-1/2$ indicates that the correlation length does
not diverge for $\phi \to \phi_c = 0$.
The threshold $\phi_c=0$ is not a true critical
point. If one insists on determining numerically the behavior of
the correlation length $\xi$ as a function of $\phi$, one finds a peak
in a position $\phi_c(N) \approx b(k_c(N))^{-1}$ but with an amplitude
that {\em decreases} with the system size ($\gamma=-1$).
In principle it is possible to define in this way an exponent
$\bar \nu = \omega/(3-\gamma_d)$, but with this value hyperscaling relation
~\eqref{eq:hyperscaling} is violated.

The rich finite size scaling phenomenology just described is summarized
in Fig.~\ref{Fig_nubar}.

The exponent $\theta$ is readily obtained from Eq.~\eqref{theta}.
In the region where hyperscaling holds $\theta=2/3$ for $\gamma_d>4$
and $\theta=(\gamma_d-2)/(\gamma_d-1)$ for $3<\gamma_d<4$.
These values coincide with those obtained using Eq.~\eqref{thetahs}.
Instead, when hyperscaling is violated, we find:
$\theta=1-(\gamma_d-3)/\omega$ for $\gamma_d>4$;
$\theta=1-1/\omega$ for $\gamma_d<4$.
These values do not coincide with those obtained from Eq.~\eqref{thetahs}:
$1/(\sigma {\bar \nu})=2(\gamma_d-3)/\omega$ for $\gamma_d>4$;
$1/(\sigma {\bar \nu})=(\gamma_d-2)/\omega$ for $\gamma_d<4$.
Table~\ref{tab:table2} presents all these results as a function
of $\omega$ for the different ranges of $\gamma_d$.

\section{Asymptotic expansions of $\psi(u)$ and $\psi'(u)$}
\label{appendix:psi_asymptotic}
In this Appendix we write the asymptotic expansions of $\psi(u^*)$ and
$\psi'(u^*)$ [see Eq.(\ref{eq:psi})] for $u^*$ close to 1 and also the expansion of
$\psi'(u)$ for  $u$ close to $u^*$. We remind that $\psi(u)=u/G_1(u)$, where
$G_1(u)=g_1(\phi u+1-\phi)$ and that $u^*$ is the solution of the
characteristic equation $\psi'(u^*)=0$.
These expansions are needed in Appendix~\ref{AppendixCalculations}
to determine the form of the cluster size distribution $n_s(\phi)$.
Indeed, in order to determine $\sigma$ we need $s_\xi$ for $|t| \ll 1$,
which can be evaluated, via Eq.~\eqref{eq:definition_s_xi},
using the expression of $\rho=\psi(u^*)$.
The condition $|t| \ll 1$ implies $u^* = 1-\eta(t)$ with $\eta \ll 1$.
Moreover, to evaluate $\tau$ we must invert Eq.~\eqref{inversion} to
find
$\epsilon=u^*-u$ as a function of $\rho-z$ and hence the exponent $a_1=a_0$,
to be inserted in Eq.~\eqref{pi_tauberian}.
In some cases this inversion process
is far from trivial and is characterized by crossovers between competing behaviors,
giving rise to preasymptotic effects in the shape of $n_s(\phi)$,
see Fig.~\ref{Figsummary}.
The main formulas we are going to use are
the Taylor expansion of $\psi(u)$ close to $u^*$, setting $\epsilon=u^*-u$,
\begin{equation}
 \psi(u) \simeq \psi(u^*) - \frac{1}{2}|\psi''(u^*)|^2 \epsilon^2,
\label{eq:taylor_psi}
 \end{equation}
where we assume $|\psi''(u^*)| \neq 0$,
and the asymptotic expansion for $u$ close to $1$, setting $\varepsilon=1-u$,
\begin{equation}
\psi(u) \simeq 1 -(1-\phi b)\varepsilon -\frac{1}{2}d \phi^2 \varepsilon^2 - C(\gamma_d-2) \phi^{\gamma_d-2} \varepsilon ^{\gamma_d-2}.
\label{eq:asymptotic_psi_general}
\end{equation}

\begin{figure}[h!]
    \includegraphics[width=0.98\columnwidth]{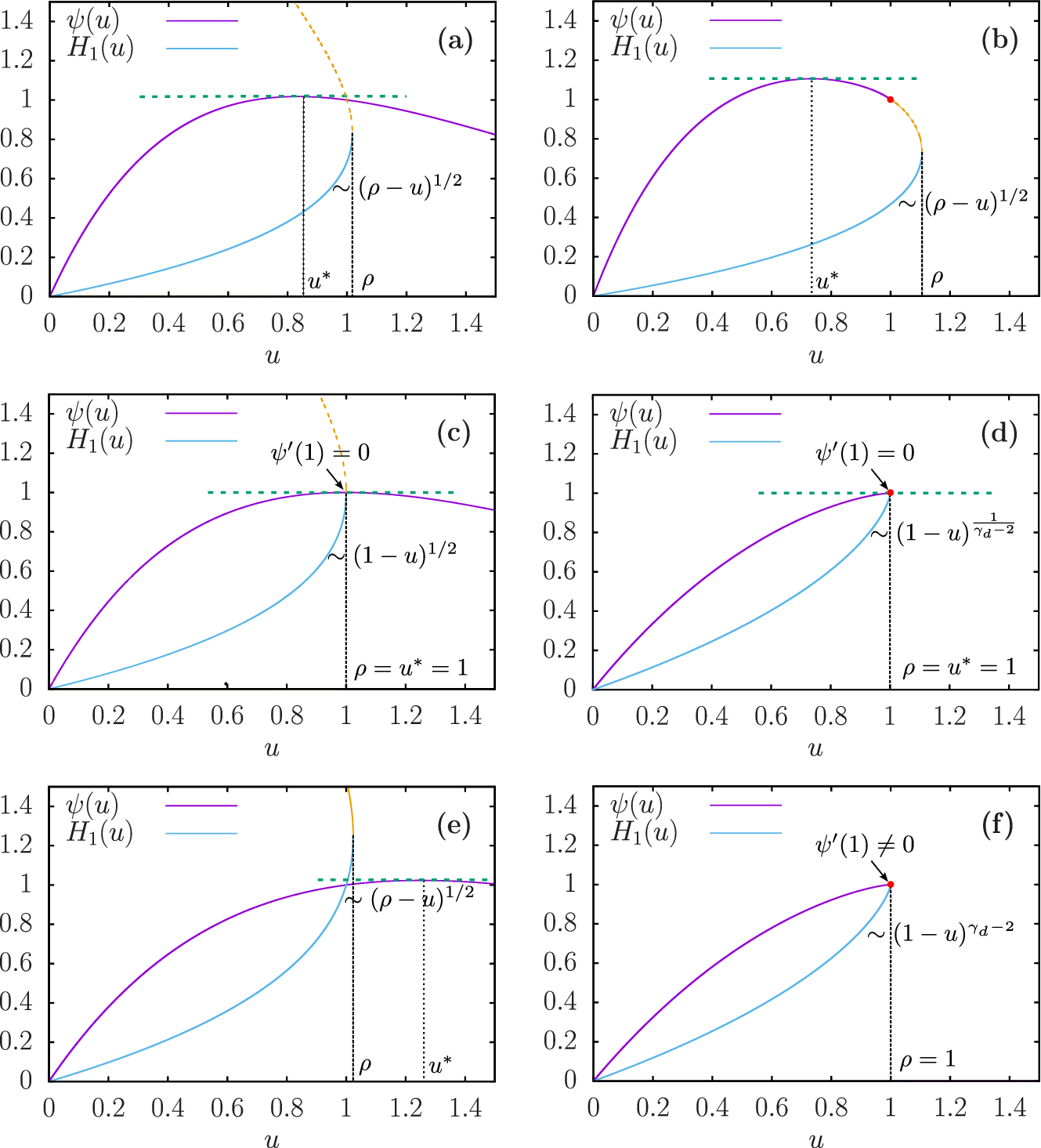}
    \caption{Various plots of the functions $\psi(u)$ (purple) and its inversion $H_1(u)$ (blue). The yellow dashed line is the non-physical part produced by the inversion of $\psi(u)$. In the different rows the cases $t>0$, $t=0$ and $t<0$, respectively, are considered. In the left column results are for homogeneous networks, in particular ER graphs, in the right column for power-law network with $\gamma=3.5$. The asymptotic scaling of $H_1$ for $\rho-u \ll 1$ is reported. Dashed lines serve as guide to the eye to indicate when the derivative of $\psi(u)$ vanishes (green dashed), in $u^*$ (black dotted), and the corresponding convergence radius $\rho$ of $H_1$ (black dashed). The red dot signals the presence of a singularity at $u=1$ in $\psi(u)$, but not necessarily in $H_1$, see panel (b). While for homogeneous networks the inversion always produces a square-root singularity, i.e. for $t>0$ (a), $t=0$ (c), $t<0$ (e), for power-law networks the inversion produces three different exponents in (b), (d), and (f) corresponding to different scaling asymptotes for $n_s(t)$. Note that the exponents in (d) and (f) look similar, but they are different, as explained in detail in Fig.~\ref{Figzoom}.}
    \label{Figinversion}
\end{figure}

\subsection{Homogeneous degree distributions}
For homogeneous distributions $G_1(z)$ is analytic and
both $\psi(u^*)$ and $\psi'(u^*)$ can be Taylor expanded
[see Fig.~\ref{Figinversion}].
In practice this amounts to use Eq.~\eqref{eq:asymptotic_psi_general} without the singular term on the r.h.s..
Evaluating it in $u^*=1-\eta(t)$, that is $\varepsilon=\eta(t)$, we have
for $|t| \ll 1$, $\eta \ll 1$,
keeping only the lowest orders in $\eta$
\begin{align}
\label{eq:psi_homogeneous}
\psi(u^*) &\simeq 1 + bt \eta,\\
\label{eq:dpsi_homogeneous}
\psi'(u^*) &\simeq -bt + d\phi_c^2 \eta .
\end{align}

The behavior of $\psi(u)$ for $u$ close to $u^*$, setting
$\epsilon=u^*-u$, is instead given by Eq.~\eqref{eq:taylor_psi}.  Note
that these expansions are valid for any $t$, both positive and
negative.

\subsection{Power-law degree distributions}
For power-law degree distributions more care must be taken.
The main problem is that $\psi(u)$ is no longer analytic in $u=1$,
because of the singularity of $G_1(u)$ in $u=1$.
Let us discuss separately the three cases
$t>0$, $t=0$, and $t<0$. In the first case, $u^*<1$ and 
$\psi(u)$ can still be Taylor expanded around $u^*$,
using Eq.~\eqref{eq:taylor_psi}.
However, as it will be shown, the
validity of such expansion breaks down as $t$ gets small.
For $t \leq 0$ instead, the singular behavior of $G_1(u)$ around
$1$ must be considered and hence Eq.~\eqref{eq:asymptotic_psi_general}
must be used.

\subsubsection{Expansions for $t>0$}
In this case, setting $u^*=1-\eta(t)$ we can write, at lowest order
in $\eta$, using Eq.~\eqref{eq:asymptotic_psi_general}
\begin{align}
\label{eq:psi_heterogeneous}
 \psi(u^*)&\simeq 1 - (1-\phi b) \eta -C(\gamma_d-2)\phi^{\gamma_d-2}\eta^{\gamma_d-2},\\
 \label{eq:dpsi_heterogeneous}
 \psi'(u^*)&\simeq (1-\phi b)+ d \phi^2 \eta +(\gamma_d-2)C(\gamma_d-2)\phi^{\gamma_d-2}\eta^{\gamma_d-3}.
\end{align}
Note that these expansions are defined only for $\eta>0$, hence for $t>0$,
in contrast to the case of homogeneous degree distributions, where $\eta<0$
for $t<0$.

Expanding $\psi(u)$ for $u$ close to $u^*$ we get again Eq.~\eqref{eq:taylor_psi} since, provided that $u^* < 1$,
we can Taylor expand $G_1(u)$ in a neighborhood of $u^*$.

However, Eq.~\eqref{eq:taylor_psi} is valid conditioned on
$\epsilon \ll 1-u^*=\eta$,
because of the presence of the singularity of $G_1(u)$ in $u=1$ (see Fig.~\ref{Figinversion}(b)).
As shown in Appendix~\ref{AppendixCalculations}, this feature leads to
preasymptotic effects in the behavior of $n_s(t)$.

\subsubsection{Expansions for $t=0$}
At criticality $u^*=1$ and the local behavior
of $\psi(u)$ for $u$ close to $u^*$ now strongly depends on
the value of $\gamma_d$.
We can no longer Taylor expand
$\psi(u)$ (see Fig.~\ref{Figinversion}(d))
and the asymptotic expansion in Eq.~\eqref{eq:asymptotic_psi_general}
must be used instead, obtaining
\begin{equation}
\psi(u) \simeq 1 -\frac{1}{2}d \phi^2 \epsilon^2 - C(\gamma_d-2) \phi^{\gamma_d-2} \epsilon^{\gamma_d-2},
\label{eq:asymptotic_psi_t=0}
\end{equation}
where $d=\langle k(k-1)(k-2) \rangle /\langle k \rangle>0$ for
$\gamma_d>4$, while $d$ is a negative constant for $\gamma_d<4$
when $\langle k^3 \rangle =\infty$.
As explicitly discussed in Appendix~\ref{AppendixCalculations},
the value of the exponent $\tau$ depends on which of the two
terms depending on $\epsilon$ in Eq.~\eqref{eq:asymptotic_psi_t=0}
is the leading order.

\subsubsection{Expansion for $t<0$}
In this case, the maximum of $\psi(u)$
is now reached in $u^*=1$, not because the first derivative
vanishes but because $\psi(u)$ is defined only for $u \leq 1$ (see Fig.~\ref{Figinversion}(f)).
Hence, setting $1-b\phi=-bt$,
\begin{equation}
 \psi(u) \simeq 1 + bt \epsilon-\frac{1}{2}d \phi^2 \epsilon^2 - C(\gamma_d-2) \phi^{\gamma_d-2} \epsilon^{\gamma_d-2},
 \label{eq:asymptotic_psi_t<0}
\end{equation}
that can be rewritten as
\begin{equation}
 1-z \simeq - bt \epsilon+\frac{1}{2}d \phi^2 \epsilon^2 + C(\gamma_d-2) \phi^{\gamma_d-2} \epsilon^{\gamma_d-2}.
 \label{eq2:asymptotic_psi_t<0}
\end{equation}

The effect of the competition between the different contributions in these expressions
on the behavior of $n_s(t)$ is
nontrivial, as reported in Appendix~\ref{AppendixCalculations}

\section{Connection with Kryven's results}
\label{Kryven}
In Ref.~\cite{kryven2017general} Kryven presented a general theory
to determine the size distribution of connected components in an
infinite network built according to the configuration model
and specified by an arbitrary degree distribution.
The expressions presented there can be used to determine
asymptotic properties of the cluster size distribution for
a percolation process on the same network by considering that,
under dilution, the tail of the degree
distribution remains a power-law with the same exponent. In
particular, if $q_k = A k^{-(\gamma_d-1)}$ is the excess degree
distribution of the network with $\phi=1$, after the dilution
$q_k(\phi) \sim A_{\phi} k^{-(\gamma_d-1)}$ for large $k$,
where $A_{\phi} = A\phi^{\gamma_d-2}$. In other words, dilution
affects the small $k$ part of the degree distribution, while the
tail remains with the same exponent, only with a different prefactor.
All the nontrivial behaviors derived by means of the generating function
approach and in particular in Fig.~\ref{Figsummary},
can be worked out from the expressions in Table II in~\cite{kryven2017general}.
In particular, it is interesting to see this connection for the case $2<\gamma_d<3$.
From Table II in~\cite{kryven2017general} and the
preasymptotic scaling of the cluster size distribution (see the
unnumbered equation below Table II) we find, using the crucial
observations that $s_{\text{Kryven}} \sim t^{\gamma_d-2}$ and
$\mu_{1,{\text{Kryven}}} = t \langle k \rangle $,
\begin{align}
 \pi_s(t)\sim \begin{cases}
               q_a t^{\tilde{a}}s^{-(\gamma_d-1)},& 1 \ll s \ll s_*,\\
               q_b t^{\tilde{b}}s^{-\frac{3}{2}}e^{-s/s_{\xi}},& s \gg s_*,
              \end{cases}
\end{align}
where $s_\xi \sim t^{-1/\sigma}$, $\sigma=(3-\gamma_d)/(\gamma_d-2)$,
$\tilde{a}=\gamma_d-1$, $\tilde{b}=1/(2\sigma)+1$, and $q_a,q_b$ are constants
independent of $t$.
This is in perfect agreement with Eq.~\eqref{eq:n_s_scalefree1}.
Similar arguments can be developed for $\gamma_d>3$, recovering the rich picture of
crossover phenomena described here.


%

\end{document}